\documentclass[%
 reprint,
 amsmath,amssymb,
pra,
floatfix,
]{revtex4-2}

\usepackage{graphicx}
\usepackage{dcolumn}
\usepackage{bm}
\usepackage{threeparttable}
\usepackage{multirow}

\newcommand{\clock}{${}^1S_0 - {}^3P_0^o$}
\newcommand{\ICL}{${}^1S_0 - {}^3P_1^o$}
\def\ai {\textit{ab initio}}
\def\CI {configuration interaction}
\def\CSF {configuration state function}
\def\ea {\textit{et al.}}
\def\HF {hyperfine}
\def\IC {intercombination}
\def\MCDHF {Multiconfiguration Dirac-Hartree-Fock}
  
\begin{document}

\preprint{APS/123-QED}

\title{MCDHF-CI calculations for Hg and Cd with estimates for unknown clock transition frequencies }

\author{Jesse S. Schelfhout} \email{jesse.schelfhout@uwa.edu.au}
\author{John J. McFerran}%
 \email{john.mcferran@uwa.edu.au}
\affiliation{%
Department of Physics, University of Western Australia, 35 Stirling Highway, 6009 Crawley, Australia
}%




\date{\today}

\begin{abstract}
By use of the \textsc{grasp2018} package we perform  Multiconfiguration Dirac-Hartree-Fock (MCDHF) calculations with configuration interaction (CI)  for the $^{1}S_{0}$ and $^{3}P_{0,1}^o$ levels in neutral cadmium and mercury. By supplying  the resultant atomic state functions to the \textsc{ris4} program, we evaluate the mass and field shift parameters for the $^{1}S_{0}-\,^{3}P_{0}^o$ (clock) and  $^{1}S_{0}-\,^{3}P_{1}^o$ (intercombination) lines. We make revised estimates of the nuclear charge parameters $\lambda^{A,A'}$ and differences in mean-square charge radii $\delta\langle r^2\rangle^{A,A'}$ for both elements and point out a discrepancy with tabulated data for Cd.  In constructing a King plot  with the Hg lines we  examine the second-order hyperfine interaction for the $^{3}P_{0,1}^o$ levels.  Isotope shifts for the clock transition have been estimated from which we predict the unknown clock line frequencies in the bosonic Hg isotopes and all the naturally occurring isotopes of Cd.
\end{abstract}

\maketitle

\section{\label{sec:1}INTRODUCTION}

Optical lattice clocks have demonstrated extraordinary levels of frequency resolution and accuracy~\cite{Bloom2014,Nemitz2016,Koller2017,McGrew2018}, as have ion-based clocks~\cite{Brewer2019,Dorscher2021}.  Such devices lend themselves well to  probing  fundamental aspects of physics~\cite{Takamoto2020,Roberts2020,BACON2021}.   Measurements undertaken with either ion clocks or lattice clocks can be used to construct King plots, where one can explore deviations from linearity~\cite{Frugiuele2017,Counts2020,Solaro2020}.  The explanations for nonlinearity is a growing topic of interest as it may yield information about phenomena lying beyond the Standard Model of particle physics~\cite{Delaunay2017,Berengut2018}.
Related studies have been carried out with ionic ytterbium~\cite{Counts2020}, neutral ytterbium~\cite{Ono2021(arXiv)}, and calcium ions~\cite{Knollmann2019,Solaro2020}, but there is the potential to extend these King plot investigations to other atomic species. 
Where optical lattice clocks can be employed, this includes neutral mercury~\cite{McFerran2012,Yamanaka2015} and cadmium~\cite{Yamaguchi2019}, which are the two  atoms of interest in the work presented here.
A key motivator is to compute the approximate optical frequencies of the \clock\ (clock) transition in all the naturally occurring isotopes of Hg and Cd.  This will aid those searching for the highly forbidden transitions (e.g. in the pursuit of a King plot analysis). 
In the case of Cd where there are, as yet, no published values for the clock transition frequencies of any of the isotopes, the best way to make these predictions is via atomic structure computations.  Our approach is as follows. (1) Compute the atomic wave functions for the lower and upper states of the clock and \ICL\ (\IC) transitions using \MCDHF\ (MCDHF) methods with \CI\ (CI). (2) Use the resultant atomic wave functions to find  hyperfine structure constants and to determine  the  mass and field shift parameters affecting isotope shifts of both the clock and the \IC\ transitions. (3) Compute the King plot slope and intercept from the mass and field shift parameters. (4)  Evaluate the isotopes shifts and absolute frequencies for the clock transition.  In the case of Hg we carry out the same procedure, but there is the added benefit of making comparisons with published values for clock transition frequencies in $^{199}$Hg and $^{201}$Hg.  Moreover, by inference from (early) experimental measurements we have a  frequency value for the clock transition in ${}^{198}$Hg~\cite{Kramida2011}, which allows us to create a King plot between \ICL\ and \clock\ lines in Hg, thereby yielding improved accuracy  for all the clock line isotope shifts.
 
 The structure of this paper runs parallel with the procedure just summarized. We begin with some details about the computational method, along with  a summary  of relevant   atomic transition data. 
  Section~\ref{sec:AbI} presents and discusses the outputs of the \ai\ atomic structure computations, which are essential for all the  calculations that follow. Section~\ref{sec:OffDiag} details the calculations associated with the second-order hyperfine (HF) interaction, which leads to a correction associated with the center of gravity of the hyperfine manifold.  In Sect.~\ref{sec:KingP} we construct a King plot between \clock\ and \ICL\ transitions in Hg, where the second order HF shift has been taken into account.  Isotope shift parameters are computed and presented in Sect.~\ref{sec:IS}, which then allow us to compute nuclear charge parameters, including differences in mean-square nuclear charge radii in Sect.~\ref{sec:Nuclear}.  The final section (\ref{sec:ClockFreqs}) presents computed isotope shifts and optical frequency predictions for  the optical clock transition in all the naturally occurring isotopes.

Within each section we  discuss Hg before Cd. This is because there is more experimental data for Hg than Cd and it therefore aids the discussion to consider Hg first. 
In most cases when we refer to ``literature values" these are experimentally determined values.

\section{\label{sec:compmethod}COMPUTATIONAL METHOD} 

The multi-configuration Dirac-Hartree-Fock (MCDHF) method with configuration interaction (CI) is used to find the atomic wavefunctions for the $^1S_0$ and ${}^3P_{0,1}^o$ levels in Hg and Cd.
From these wavefunctions we determine the diagonal and off-diagonal hyperfine interaction constants, and   mass- and field-shift parameters related to isotope shifts.  The \textsc{grasp2018} (General Relativistic Atomic Structure Package 2018) package \cite{Froese_Fischer2018} is used to perform MCDHF-CI computations for atomic state functions and the \textsc{ris4} (Relativistic Isotope Shifts 4) program \cite{Ekman2019} is used to calculate the isotope shift parameters from them.  The theoretical background for the methods is  detailed in Refs.~\cite{Froese_Fischer2016,Grant2007pp235-403}. The computations were performed using 34 cores on one node of Kaya, a small supercomputer administered by the UWA High Performance Computing team. 

The  MCDHF-CI calculations for Hg and Cd follow a similar approach used for Yb in our previous work \cite{Schelfhout2021(2)}. For Cd, with the ground state valence electrons filling the $5s$ orbital, an additional correlation layer is required to extend the active space to $12s,12p,12d,12f$ compared to Yb and Hg, for which the ground state valence electrons fill the $6s$ orbital. The main point of difference concerns the multiconfiguration correlation model adopted for Hg and  Cd. The outputs produced by starting from a single-reference (SR) configuration (for the zeroth order wave function) are found to be in better agreement with experimental values than results from a multi-reference (MR) set  as used in  \cite{Schelfhout2021(2)} (but more so for Cd). This difference could be due to the presence of a closed $d$-shell just under the valence shell in Cd and Hg that is absent in Yb --- a hypothesis that could be tested by performing computations using both SR and MR set approaches for nobellium and copernicium (but beyond the scope of the present work).

Calculated results are compared with experimental values where possible to gauge their error. In addition, we quantify computational errors by examining how mass- and field-shift parameters converge as correlation layers are added to the active space and as core-correlation deepens.  This is described in some detail in Appendix~\ref{app:Uncertainties}.  The centroid of the hyperfine manifold is used for fermionic isotopes to make a comparison with bosonic isotopes. Additionally, the off-diagonal terms in the hyperfine interaction lead to shifts to the $F=I$ levels for the ${}^3P_{0,1}^o$ states; these shifts affect the centroids and so experimental measurements have been offset to account for them. The comparisons between computed isotope shifts and experimentally determined isotope shifts take into account the second-order hyperfine perturbation.

\begin{figure}[h]
    \centering
    \includegraphics[width=\columnwidth]{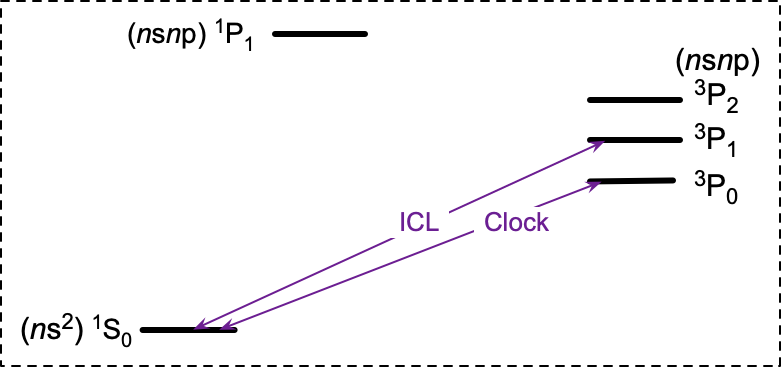} 
    \caption{Low lying energy level structure in divalent atoms with $LS$-coupling notation. ICL, intercombination line.}
    \label{fig:energylevels} 
\end{figure}
%


Mercury and cadmium are group IIb atoms, and like all divalent atoms have a low lying energy level structure similar to that in Fig.~\ref{fig:energylevels}, where the clock and intercombination lines are denoted. The ground state principal quantum number $n=5\, (6)$  for Cd (Hg).  
Mercury ($Z=80$) has seven stable isotopes, including five bosonic isotopes. Some properties for the isotopes (in the nuclear ground state) are presented in Table \ref{tab:Hg nuclear properties} of Appendix~\ref{app:a}. The two fermionic isotopes are $^{199}$Hg and $^{201}$Hg with nuclear spin $I=1/2$ and $3/2$, respectively. The ground state electron configuration for Hg is [Xe]$4f^{14}5d^{10}6s^2$. 
A number of absolute transition frequency values exist in the literature for the clock line and intercombination lines (ICL)  in Hg isotopes. These values are presented in Table \ref{tab:Hg absolute values}. They are used as a point of comparison for the \textit{ab initio} calculations, and in the calculation of  the frequencies for the unexplored clock transitions (we have not found absolute values of $\nu_\mathrm{ICL}$ for $^{199}$Hg and $^{201}$Hg, but there are hyperfine level separations, which are used later in Sec.~\ref{sec:Centroid}).  

\begin{table}[h]
    \centering
    \caption{Literature values for the absolute transition frequencies of the \clock\ clock and \ICL\ intercombination lines in isotopes of Hg \textsc{i}.}
     \begin{ruledtabular}
    \begin{tabular}{lcdc} 
        Isotope & Transition & \multicolumn{1}{c}{$\nu$(GHz)} & Ref.\\ \hline
       198 & clock & 1128575.955(11) & \cite{Kramida2011}\\
       199 & clock & 1128575.29080815462(41) & \cite{Tyumenev2016}\\
        201 & clock & 1128569.5611396(53) & \cite{Petersen2008}\\
        198 & ICL & 1181555.77854(27) & \cite{Witkowski2019}\\
        200 & ICL & 1181550.97226(20) & \cite{Witkowski2019}\\
        202& ICL & 1181545.67685(11) & \cite{Witkowski2019}\\
        204 & ICL & 1181540.46653(14) & \cite{Witkowski2019}\\ 
    \end{tabular}
    \end{ruledtabular}
    \label{tab:Hg absolute values}
\end{table} 

For Hg, the starting point single-reference configuration  for ${}^1S_0$ is $\{6s^2\}$ and for ${}^3P_{0,1}^o$ it is $\{6s6p\}$. 
The core orbitals made available for substitution, to account for core-valence correlation, are $\{5d,5p,5s,4f\}$. A minor issue arises when estimating the $f$-orbitals for the correlation layers in Hg using the Thomas-Fermi potential. This issue is overcome by using screened hydrogenic estimates where necessary. A further issue arises with the convergence of the $5f_{\pm}$-orbitals in the first correlation layer for $^1S_0$. 
A solution was found that allowed the $5f_{\pm}$-orbitals to converge according to the convergence criteria for the self-consistent field procedure (rather than by hitting the iteration limit). The $5f_{\pm}$-orbitals were computed on their own
with a reduced rate-of-change to the \CSF\ (CSF) expansion coefficients per iteration.
This was followed by the remainder of the orbitals for the first correlation layer being computed without the $5f_{\pm}$-orbitals and with standard rate-of-change to the  CSF expansion coefficients.
All of the orbitals for the correlation layer were then computed together with
a reduced rate-of-change to the CSF expansion coefficients. 

Cadmium ($Z=48$) has eight stable isotopes, including six bosonic isotopes. Some properties for the  (nuclear ground state) isotopes are presented in Table \ref{tab:Cd nuclear properties} of Appendix~\ref{app:a}. Both of the fermionic isotopes, $^{111}$Cd and $^{113}$Cd, have nuclear spin $I=1/2$ and are  good candidates for optical lattice clocks~\cite{Yamaguchi2019}.
Cd is the analogue element to Hg with one fewer layer of each of the orbitals. Accordingly, an additional correlation layer is required to bring Cd up to an active space of $12s,12p,12d,12f$ compared with Yb and Hg. The ground state electron configuration for Cd  is
[Kr]$4d^{10}5s^2$.
The SR  for  ${}^1S_0$  is $\{5s^2\}$ and for ${}^3P_{0,1}^o$ it is $\{5s5p\}$. 
The expansion of the active core to core-valence correlation is discussed in Sec.~\ref{sec:AbI} and Appendix~\ref{app:Uncertainties}.

Absolute  frequency measurements for  the \ICL\ transition in $^{114}$Cd are available in Ref.~\cite{Burns1956} (Burns and Adams), however, direct measurements for the clock transition in Cd  are not (yet) found in the literature. Instead, the difference between ${}^3P_0^o - {}^3S_1$ and ${}^3P_1^o- {}^3S_1$ transitions in  Ref.~\cite{Burns1956} can be
used to infer the frequency of the clock transition.
The frequencies for the clock and intercombination lines in  ${}^{114}\mathrm{Cd}$ are found to be, $\nu_{\mathrm{clock}}({}^{114}\mathrm{Cd})=902794.62(6)$\,GHz and $\nu_{\mathrm{ICL}}({}^{114}\mathrm{Cd})=919046.19(6)$\,GHz, respectively.
These can be compared with the \textit{ab initio} calculations and used later to  calculate absolute frequencies of  the unexplored clock transition frequencies in Cd (Sec.~\ref{sec:ClockFreqs}). 

\section{\label{sec:AbI}Ab initio transition frequencies and isotope shifts}

The atomic structure software package \textsc{grasp2018} outputs energy levels of the atomic state functions for each atomic state separately (in Hartree energy units $E_h$). The transition frequencies are calculated from the differences between the energies of the states involved in the transitions and are converted into SI units using the factor $2cR_{\infty}=6.579 683 920 502(13) \times 10^{15}\,{E_h}^{-1}\,\mathrm{Hz}$ \cite{Tiesinga2021}. These \textit{ab initio} transition frequencies are calculated for each isotope along with isotope shifts relative to ${}^{198}$Hg for the \clock\ clock transition  (Table \ref{tab:Hg clock ab initio energies}) and for the \ICL\ \IC\ line (ICL)  (Table \ref{tab:Hg ICL ab initio energies}).

\begin{table}[h]
    \centering
    \caption{\textit{Ab initio} calculations for the energy level difference between ${}^3P_0^o$  and ${}^1S_0$, and the associated isotope shifts  in isotopes of Hg \textsc{i}.}
    \begin{ruledtabular}
    \begin{tabular}{ccD{.}{.}{3.3}} 
        A & $\Delta E$ for \clock\  (GHz) & \multicolumn{1}{l}{$\delta\nu_{\mathrm{clock}}^{A,198}$\,(GHz)} \\ \hline
        196 & 1128734.558 & 4.370\\
        198 & 1128730.188 & 0\\
        199 & 1128729.585 & -0.603\\
        200 & 1128725.166 & -5.022\\
        201 & 1128723.499 & -6.689\\
        202 & 1128719.682 & -10.506\\
        204 & 1128714.245 & -15.943\\  
    \end{tabular}
     \end{ruledtabular}
    \label{tab:Hg clock ab initio energies}
\end{table}

For the clock transition, the absolute transition frequencies presented in Table \ref{tab:Hg clock ab initio energies} are at $0.02\%$ difference with the  values in Table \ref{tab:Hg absolute values}. Thus, there is excellent agreement with the experimental values and they are closer than  values computed previously~\cite{Angstmann2004,Dinh2008,Gogyan2021,Li2007}.
This is also more than an order of magnitude more accurate than the results for Yb \cite{Schelfhout2021(2)}. The isotope shifts $\delta\nu_{\mathrm{clock}}^{199,198}$ and $\delta\nu_{\mathrm{clock}}^{201,198}$ are at errors of about -9\% and 5\%, respectively, in comparison to the (second-order corrected) experimental values in Table \ref{tab:corrected Hg line shifts}. These \textit{ab initio} isotope shifts for Hg are also more accurate than those for Yb \cite{Schelfhout2021(2)}.

\begin{table}[h]
    \centering
    \caption{\textit{Ab initio} calculations for the energy level difference between ${}^3P_1^o$  and ${}^1S_0$, and the associated isotope shifts  in isotopes of Hg \textsc{i}.}
     \begin{ruledtabular}
    \begin{tabular}{ccD{.}{.}{3.3}} 
        A & $\Delta E$ for \ICL\  (GHz) & \multicolumn{1}{l}{$\delta\nu_{\mathrm{ICL}}^{A,198}$\,(GHz)} \\ \hline
        196 & 1179728.244 & 4.406\\
        198 & 1179723.838 & 0\\
        199 & 1179723.228 & -0.610\\
        200 & 1179718.784 & -5.054\\
        201 & 1179717.102 & -6.736\\
        202 & 1179713.259 & -10.579\\
        204 & 1179707.778 & -16.060\\ 
    \end{tabular}
    \end{ruledtabular}
    \label{tab:Hg ICL ab initio energies}
\end{table}

For the ICL transition, the absolute transition frequencies presented in Table \ref{tab:Hg ICL ab initio energies} are at $-0.2\%$ difference with the values in Table \ref{tab:Hg absolute values}. This represents good agreement with experiment and an improvement on the calculated values in Refs.~\cite{Angstmann2004,Glowacki2003,Gogyan2021,Li2007,Migdalek1985,Yu2007}, but  less accurate than values in Ref.~\cite{Dinh2008}. Whilst the accuracy is an order of magnitude poorer than for the clock transition, this is approximately a factor of four more accurate than the equivalent results for Yb \cite{Schelfhout2021(2)}. The isotope shifts are at errors of about 5\% in comparison to the experimental values in Refs.~\cite{Witkowski2019,Fricke2004_Hg}. 
This discrepancy  could be due to an imperfect electron correlation model or due to the impact of nuclear deformation that is ignored in these computations.

For cadmium, the \textit{ab initio} transition frequencies are calculated for each isotope, and the isotope shifts are evaluated relative to ${}^{114}$Cd for the \clock\ clock transition (Table \ref{tab:Cd clock ab initio energies}) and for the \ICL\ ICL (Table \ref{tab:Cd ICL ab initio energies}).
\begin{table}[h]
    \centering
    \caption{\textit{Ab initio} calculations for the energy level difference between ${}^3P_0^o$  and ${}^1S_0$, and the associated isotope shifts in isotopes of Cd \textsc{i}.}
     \begin{ruledtabular}
    \begin{tabular}{ccd} 
        A & $\Delta E$ for \clock\   (GHz) & \multicolumn{1}{l}{$\delta\nu_{\mathrm{clock}}^{A,114}$\,(MHz)} \\ \hline
        106 & 895992.29560 & 1542.80\\
        108 & 895991.86167 & 1108.87\\
        110 & 895991.43162 & 678.82\\
        111 & 895991.26930 & 516.50\\
        112 & 895991.02632 & 273.52\\
        113 & 895990.90407 & 151.27\\
        114 & 895990.75280 & 0\\
        116 & 895990.57475 & -178.05\\ 
    \end{tabular}
     \end{ruledtabular}
    \label{tab:Cd clock ab initio energies}
\end{table}

For the clock transition, the absolute transition frequency for ${}^{114}$Cd presented in Table \ref{tab:Cd clock ab initio energies} has a $-0.8\%$ difference from the Burns and Adams value  stated above. 
We are not aware of any other calculated values for the clock transition frequency. Although the error is more than an order of magnitude larger than for Hg,  as was the case in Yb~\cite{Schelfhout2021(2)}, 
it indicates that the atomic wave functions are sufficiently accurate to derive further atomic and nuclear parameters.
No comparison can be made with experimental values for the isotope shifts to gauge their error; however, the error is expected to be similar to that found for the ICL transition.

\begin{table}[h]
    \centering
    \caption{\textit{Ab initio} calculations for the energy level difference between ${}^3P_1^o$  and ${}^1S_0$, and the associated isotope shifts in isotopes of Cd \textsc{i}.}
     \begin{ruledtabular}
    \begin{tabular}{ccd} 
        A & $\Delta E$ for \ICL\  (GHz) & \multicolumn{1}{l}{$\delta\nu_{\mathrm{ICL}}^{A,114}$\,(MHz)} \\ \hline
        106 & 910815.77401 & 1566.36\\
        108 & 910815.33356 & 1125.91\\
        110 & 910814.89752 & 689.87\\
        111 & 910814.73218 & 524.53\\
        112 & 910814.48616 & 278.51\\
        113 & 910814.36108 & 153.43\\
        114 & 910814.20765 & 0\\
        116 & 910814.02427 & -183.38\\ 
    \end{tabular}
     \end{ruledtabular}
    \label{tab:Cd ICL ab initio energies}
\end{table}

For the ICL transition, the absolute transition frequency for ${}^{114}$Cd presented in Table \ref{tab:Cd ICL ab initio energies} has a $-0.9\%$ difference from the Burns and Adams value. 
This is a fair agreement with experiment and improves on estimates computed in Refs.~\cite{Glowacki2003,Migdalek1988,Yu2007}, but is slightly less accurate than that of Ref.~\cite{Biemont2000}. 
The isotope shifts are at errors of about $-20\%$ to $-30\%$ in comparison to the  values in Ref.~\cite{Fricke2004_Cd}, except for $\delta\nu_{\mathrm{ICL}}^{113,114}$ which is at more than $-50\%$ error. These \textit{ab initio} isotope shifts for Cd are  less accurate than those for Hg.
With an additional core orbital ($4d4p4s3d$) opened to core-valence correlation,  the difference for $\delta\nu_{\mathrm{ICL}}^{113,114}$ reduced to 45\%, but that for $\delta\nu_{\mathrm{ICL}}^{116,114}$ increased to 85\%, 
and so the deeper core computations were not pursued (the field shift parameter $F$, discussed below, also became more disparate with previous estimates as the core-valence correlation deepened).  

The weaker agreement with experimental values for the \textit{ab initio} isotope shifts of Cd in comparison to Yb and Hg could be an indicator of the suitability of this computational method for heavier atoms compared to lighter atoms. Relativistic effects become more important with increasing $Z$. 
A more suitable approach may be to use a non-relativistic multiconfiguration Hartree-Fock (MCHF) computation with relativistic effects accounted for through the Breit-Pauli corrections  (better suited for light and near neutral systems with smaller relativistic effects~\cite{Froese_Fischer2016,Froese_Fischer2018}).

\section{\label{sec:OffDiag}Diagonal and off-diagonal hyperfine constants} 

The fermionic isotopes of Cd and Hg experience the hyperfine interaction. To second-order, the off-diagonal hyperfine interaction causes a shift to the $F=I$ levels in the hyperfine manifolds, which Kischkel~\ea~\cite{Kischkel1991} denote as $\Delta E_F^{(2)}$.  Here we will refer to it as  $\Delta \nu_F^{(2)}$, in line with our use of hertz for units.  The second-order HF perturbation is also described (or summarized) in Refs.~\cite{Wakasugi1990,Jonsson1996,Grant2007}. 
These shifts are dependent upon the hyperfine interaction constants, which we have  calculated from the atomic state functions using the \textsc{rhfs} program~\cite{Jonsson1996,Froese_Fischer2018}. 
Our computed hyperfine constants appear in Table \ref{tab:Hg hfs constants}. The uncertainties have been estimated by comparing the calculated values with the experimental values from \cite{Bonn1976}. The errors in the diagonal magnetic dipole constants $A({}^3P_1^o)$ are 6\% and this uncertainty is extended to the off-diagonal magnetic dipole constants $A({}^3P_0^o,{}^3P_1^o)$, while the error in the electric quadrupole constant, $B({}^3P_1^o)$, for ${}^{201}$Hg is 0.4\%.

\begin{table}[h]
    \centering
    \caption{Calculated hyperfine interaction constants for Hg in units of GHz.} 
    \begin{ruledtabular}
    \begin{tabular}{cccc} 
        $A$ & $A({}^3P_1^o)$ & $A({}^3P_0^o,{}^3P_1^o)$ & $B({}^3P_1^o)$ \\ \hline
        199 & 15.59(94) & 14.42(87) & \\
        201 & -5.75(35) & -5.32(32) & -0.2812(12)\\ 
    \end{tabular}
    \end{ruledtabular}
    \label{tab:Hg hfs constants}
\end{table}

Calculation of the shifts to the $F=I$ levels requires the fine structure interval, for which the difference between the ICL and clock transition frequencies of ${}^{198}$Hg in Table \ref{tab:Hg absolute values} has been used [giving a value of 52979.823(11)\,GHz]. The calculation is as in Ref.~\cite{Schelfhout2021(2)} for Yb~\cite{comment1}, and the values are presented in Table \ref{tab:Hg Delta E}. The notation $\Delta \nu_F^{(2)}({}^3P_0^o,{}^3P_1^o)$ implies that the magnitude of the shift is the same for both the ${}^3P_0^o$ and ${}^3P_1^o$ levels. The shift is negative (positive) for ${}^3P_0^o$ (${}^3P_1^o$).   The uncertainties are propagated from those in Table \ref{tab:Hg hfs constants}. For comparison, a value for the  $\Delta \nu_F^{(2)}$ shift in ${}^{199}$Hg ($I=1/2$) is determined using the experimental hyperfine structure splitting $\Delta\nu_{\mathrm{hfs}}$ presented in Ref.~\cite{Stager1963} (that leads to an effective HF structure constant, following~\cite{Wakasugi1990}) and the diagonal HF interaction constant $A$ that includes second-order corrections~\cite{Bonn1976}, according to the formula 
\begin{equation}\label{eq:Delta E}
    \Delta \nu_F^{(2)} = \frac{3}{2} A - \Delta\nu_{\mathrm{hfs}}.
\end{equation}
A discussion of Eq.~(\ref{eq:Delta E}) appears in Appendix \ref{app:centroid}. Our value for ${}^{199}$Hg is consistent with that calculated using literature values within the uncertainties (see Table~\ref{tab:Hg Delta E}). Given that the literature value accounts for shifts due to other levels, this consistency suggests the influence of levels other than ${}^3P_0^o$ on  ${}^3P_1^o$  is less significant. 
We could not find published data to make a similar comparison for ${}^{201}$Hg. 
While these second order shifts are small compared to the uncertainties for the isotope shifts evaluated below, in time, when  the isotope shifts are measured carefully, this higher order shift needs to be taken into account.  

\begin{table}[h]
    \centering
    \caption{Corrections $\Delta \nu_F^{(2)}({}^3P_0^o,{}^3P_1^o)$  to the $F=I$ hyperfine levels in Hg due to second-order off-diagonal hyperfine interaction. The literature value is calculated using experimental results from Refs. \cite{Stager1963,Bonn1976} as described in the text.}
    \begin{ruledtabular}
    \begin{tabular}{ccc} 
       {$A$} & \multicolumn{2}{c}{$\Delta \nu_F^{(2)}({}^3P_0^o,{}^3P_1^o)$\,(MHz)}\\ 
        & This work & Literature\\ \hline
        199 & 2.94(36) & 2.65(8)\\
        201 & 2.00(25) \\ 
    \end{tabular}
    \end{ruledtabular}
    \label{tab:Hg Delta E}
\end{table}

For Cd, our calculated hyperfine interaction constants  are presented in Table \ref{tab:Cd hfs constants}. The uncertainties have been estimated by comparing the calculated values with experimental values derived from the literature according to
\begin{align}\label{eq:mumbo-jumbo}
    A({}^3P_1) = & \frac{1}{4}(2c_2^2 - c_1^2)a_s + \frac{5}{4}c_1^2a_{3/2} \nonumber\\ 
  & + \frac{1}{2}c_2^2a_{1/2} - \frac{5\sqrt{2}}{16}c_1c_2\xi a_{3/2},
\end{align}
as described in Refs.~\cite{Garstang1962,Breit1933}. The $c$ coefficients relate $jj$-coupled states to intermediate coupled states and depend on mixing coefficients, $\alpha$, $\beta$, where  $|^3P_1\rangle = \alpha |^3P_1\rangle_\mathrm{LS} +\beta|^1P_1\rangle_\mathrm{LS}$. The $a$ coefficients are hyperfine coefficients in the $|j_1 j_2 J m_J\rangle$ basis. The values for the terms on the right-hand side of (\ref{eq:mumbo-jumbo}) are taken from \cite{Thaddeus1962} except for $\xi$, which is from \cite{Garstang1962}. These give  values of $A({}^3P_1^o) = -4.124(32)$\,GHz for ${}^{111}$Cd and $A({}^3P_1^o) = -4.314(34)$\,GHz for ${}^{113}$Cd.  The difference between these values and our computed values is  0.04\%, for both isotopes. We therefore assign errors in our  $A({}^3P_1^o)$ constants to be  0.04\% and this  is extended to the off-diagonal magnetic dipole constants $A({}^3P_0^o,{}^3P_1^o)$.

\begin{table}[h]
    \centering
    \caption{Our calculated hyperfine interaction constants for Cd.} 
    \begin{ruledtabular}
    \begin{tabular}{ccc}
        $A$ & $A({}^3P_1^o)$\,(GHz) & $A({}^3P_0^o,{}^3P_1^o)$\,(GHz)\\ \hline
        111 & -4.1256(17) & -4.4807(18) \\
        113 & -4.3156(18) & -4.6869(19) \\ 
    \end{tabular}
    \end{ruledtabular}
    \label{tab:Cd hfs constants}
\end{table}

Calculation of the shifts to the $F=I$ levels follows the same procedure as for Hg, 
except no literature values exist or were able to be calculated from published data for comparison. The values are presented in Table \ref{tab:Cd Delta E} and uncertainties are propagated from those in Table \ref{tab:Cd hfs constants}.

\begin{table}[h]
    \centering
    \caption{Corrections $\Delta \nu_F^{(2)}({}^3P_0^o,{}^3P_1^o)$ to the $F=I$ hyperfine levels in Cd due to second-order off-diagonal hyperfine interaction.}
    \begin{ruledtabular}
    \begin{tabular}{cd}  
        $A$ & \multicolumn{1}{c}{$\Delta \nu_F^{(2)}({}^3P_0^o,{}^3P_1^o)$\,(kHz)} \\ \hline
        111 & 926.53(75) \\
        113 & 1013.76(82) \\ 
    \end{tabular}
    \end{ruledtabular}
    \label{tab:Cd Delta E}
\end{table}

\section{\label{sec:Centroid} Centroid frequencies}

The second-order off-diagonal shifts to the $F=I$ levels 
for the fermionic isotopes (calculated in Section \ref{sec:OffDiag}) lead to shifts to the centroids (or centers of gravity) of the hyperfine manifolds. For the mutual influence of the ${}^3P_{0,1}^o$ states, the observed $F=I$ levels  (which includes the $\mathcal{O}^{(2)}$  shifts) are at higher energy for  ${}^3P_1^o$  by $\Delta \nu_F^{(2)}$, and at lower energy for  ${}^3P_0^o$  by the same amount. These shifts are considered relative to a theoretical atomic system which does not experience an off-diagonal hyperfine interaction in the basis of fine-structure eigenstates (but does experience a diagonal interaction). The centers of gravity for the fermionic isotopes are used in comparison with the bosonic isotopes and these higher-order shifts need to be accounted for. The experimental isotope shifts ($\delta\nu^{A,198}$) involving the fermionic isotopes have been adjusted to remove the influence of the off-diagonal hyperfine interaction. These  values for Hg are presented in Table \ref{tab:corrected Hg line shifts} together with experimental isotope shifts involving only the bosonic isotopes.
For the purposes of this paper we ``adjust" or “correct” the experimental values for the second order \HF\ interaction because we wish to use the linearity of a King plot to make predictions.  However, we understand that from an experimenter’s point of view it should be the computational results that are corrected and not the reverse.  

The uncertainties for $\delta\nu^{A,198}$ for the fermionic isotopes are currently limited by the precision of the isotope shifts rather than the uncertainties associated with the correction. This is in contrast to the situation for corrections to  $\delta\nu_{\mathrm{clock}}^{199,201}$, which is not used here. 

\begin{table}[h]
    \centering
    \caption{Literature isotope shifts for the \clock\ clock and \ICL\ intercombination lines in Hg \textsc{i}. $\delta\nu^{A,198} = \nu^A - \nu^{198}$. The  centroids for the hyperfine manifolds in fermionic isotopes have been corrected as outlined in the text.}
    \begin{ruledtabular}
    \begin{tabular}{ccD{.}{.}{3.9}c}   
        $A$ & Transition & \multicolumn{1}{c}{$\delta\nu^{A,198}$\,(GHz)} & Refs.\\ \hline
        199 & clock & -0.662(11) & \cite{Tyumenev2016,Kramida2011}\\
        201 & clock & -6.392(11) & \cite{Petersen2008,Kramida2011}\\
        196 & ICL & 4.199(4) & \cite{Fricke2004_Hg}\\
        199 & ICL & -0.6524(63) & \cite{Fricke2004_Hg}\\
        200 & ICL & -4.80628(33) & \cite{Witkowski2019}\\
        201 & ICL & -6.4101(57) & \cite{Fricke2004_Hg}\\
        202 & ICL & -10.10174(28) & \cite{Witkowski2019}\\
        204 & ICL & -15.31202(30) & \cite{Witkowski2019}\\ 
    \end{tabular}
    \end{ruledtabular}
    \label{tab:corrected Hg line shifts}
\end{table}

Following Kischkel~\ea~\cite{Kischkel1991}, a consistency check of the second order HF corrections can be made when there are at least two odd isotopes.  Two odd isotopes (labelled $f$ and $f'$)  have different isotope shifts with respect to a bosonic isotope (labeled $b$), which following from our adopted notation could be written as $\delta\nu^{f,b}$ and  $\delta\nu^{f',b}$. 
Each experiences a second order HF correction affecting the center of gravity, which we denote as $\Delta\nu^{(2)}_{f,b}$, in the case of isotope $f$.
The ratio of these second order HF corrections can be expressed as
\begin{equation}\label{eq:quotient}
    \frac{ \Delta\nu^{(2)}_{f,b}}{\Delta\nu^{(2)}_{f',b}} = \frac{(-1)^{2I_f+1}\mu_f^2(I_f+1)I_{f'}}{(-1)^{2I_{f'}+1}\mu_{f'}^2(I_{f'}+1)I_f},
\end{equation}
where $I$ is the nuclear spin and $\mu$ is the nuclear magnetic moment.
For Hg, the right-hand side of (\ref{eq:quotient}) evaluates to $1.46775(1)$ (with ${}^{199}$Hg in the numerator) using values from Table \ref{tab:Hg nuclear properties}. For the ${}^3P_0^o$ state, the adjustments to the isotope shifts are as in Table \ref{tab:Hg Delta E} and the left-hand side of (\ref{eq:quotient}) evaluates to $1.47(26)$, showing good agreement. For the ${}^3P_1^o$ state the adjusted centroids are calculated directly from shifts to the hyperfine structure levels of \cite{Stager1963} and so (\ref{eq:quotient}) is not evaluated. The consistency between the ratios for the clock excited state supports the accuracy of the calculated shifts.

In the case of Cd there are no experimental isotope shifts of the clock transition to adjust, and the corrections for the ICL transition are equal to one-third of the shifts presented in Table \ref{tab:Cd Delta E} (see Appendix \ref{app:centroid}) thus fall below the level prescribed by experimental uncertainty of the isotope shifts from \cite{Fricke2004_Cd}. These experimental isotope shifts for the ICL transition are presented in Table \ref{tab:corrected Cd line shifts}.

\begin{table}[h]
    \centering
    \caption{Literature isotope shifts for the \ICL\ intercombination line in Cd \textsc{i} from \cite{Fricke2004_Cd}. $\delta\nu^{A,114} = \nu^A - \nu^{114}$. The adjustments to the centroids for the hyperfine manifolds of fermionic isotopes are below the level of experimental uncertainty.}
    \begin{threeparttable}
    \begin{ruledtabular}
    \begin{tabular}{cD{,}{}{4.4}} 
        $A$ & \multicolumn{1}{c}{$\delta\nu_{\mathrm{ICL}}^{A,114}$\,(MHz)} \\ \hline
        106 & 1943,(32)\\
        108 & 1385,(42)\\
        110 & 893,(19)\\
        111 & 839,(49)\\
        112 & 405,(12)\\
        113 & 333,(41) \tnote{a}\\
        116 & -279,(12)\\ 
    \end{tabular}
    \end{ruledtabular}
    \begin{tablenotes}\footnotesize
    \item[a]{Consistent with the original reference of \cite{Kelly1959}.}
    \end{tablenotes}
    \end{threeparttable}
    \label{tab:corrected Cd line shifts}
\end{table} 

For Cd, the two fermionic isotopes have the same 1/2 nuclear spin, hence the quotient of Eq.~(\ref{eq:quotient}) becomes the
 ratio of squared nuclear magnetic moments. With ${}^{111}$Cd in the numerator, this evaluates to $0.913833(4)$ using the values in Table \ref{tab:Cd nuclear properties}. For the ${}^3P_0^o$ state, the second-order corrections to the isotope shifts are as in Table \ref{tab:Cd Delta E} and the ratio of the centroid corrections  (LHS of Eq.~\ref{eq:quotient}) evaluates to $0.9140(11)$, consistent with the RHS of Eq.~\ref{eq:quotient}. This will also equal the ratio for the ${}^3P_1^o$ state since the adjustment to the center of gravity is equal to one-third of the shift to the $F=I$ level for $I=1/2$ and both fermionic isotopes have $I=1/2$.

\section{\label{sec:KingP}King plot}

\begin{figure}[h]
    \centering
    \includegraphics[width=1.01\columnwidth]{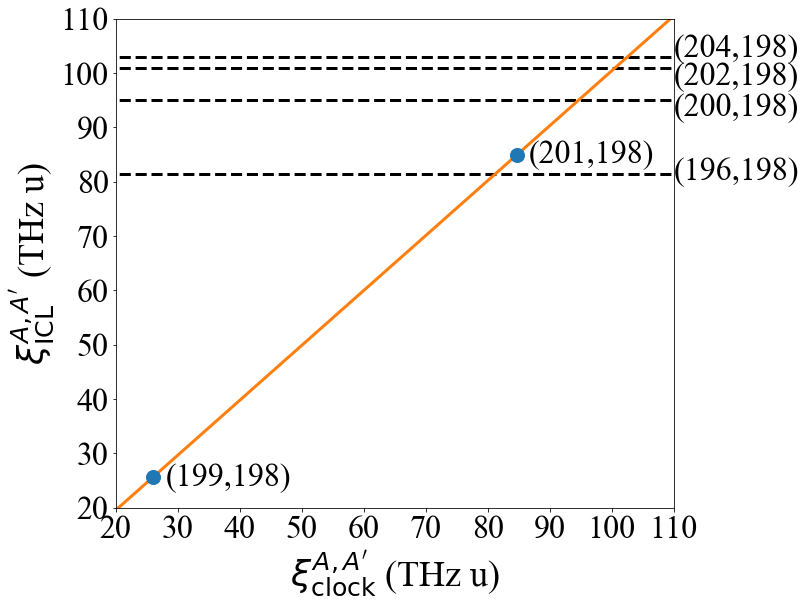} 
    \caption{King plot for clock and intercombination lines of mercury. Blue circles represent isotope pairs with available data for the clock transition frequency. Dashed black lines represent isotope pairs without clock transition measurements (only ICL data). The solid orange line is the King linearity relationship. Pairs of integers in parentheses indicate isotope pairs. Error bars are not visible at the scale of this plot.}
    \label{fig:Hg King plot}  
\end{figure}

For an electronic transition, labelled by $a$, the isotope shift between isotopes $A$ and $A'$ is split into a mass shift, due to the difference in nuclear recoil, and a field shift, due to the difference in the nuclear electrostatic potential, and is given by
\begin{equation}\label{eq:isotope shift}
    \delta\nu_{a}^{A,A'} = \nu_{a}^{A} - \nu_{a}^{A'} = \mu^{A,A'} K_{a} + \lambda^{A,A'} F_{a},
\end{equation}
where $\mu^{A,A'} = m_{A}^{-1}-m_{A'}^{-1}$ is the nuclear mass shift term, $K_{a}$ is the electronic mass shift term, $F_{a}$ is the electronic field shift term \cite{King1984}, and 
\begin{equation}\label{eq:lambda}
    \lambda^{A,A'} = \delta\langle r^2 \rangle^{A,A'} \left( 1 + \sum_{n=1}^{\infty} C_n \frac{\delta\langle r^{2n+2} \rangle^{A,A'}}{\delta\langle r^2 \rangle^{A,A'}} \right)
\end{equation}
is the nuclear charge parameter. The $C_n$ terms in Eq.~(\ref{eq:lambda}) are Seltzer's coefficients \cite{Seltzer1969} and $\delta\langle r^2\rangle^{A,A'}$ is the difference in mean-square charge radii between the nuclei of isotopes $A$ and $A'$. Accordingly, isotope shifts are sensitive to the size and shape of the atomic nucleus and optical isotope shifts are used to determine differences in mean-square nuclear charge radii \cite{Angeli2013}. Equation (\ref{eq:isotope shift}) is a good approximation, but neglects higher-order contributions due to differences in nuclear shape, such as the quadratic field shift and differential nuclear deformation \cite{Allehabi2021}.
Following King~\cite{King1963}, the isotope shift in Eq.~\ref{eq:King} can be mass-scaled such that $\xi_{a}^{A,A'} = \delta\nu_{a}^{A,A'} / \mu^{A,A'}$, from which one can deduce a straightforward relationship between  isotope shifts of different transitions, namely,
\begin{equation}\label{eq:King}
    \xi_{b}^{A,A'} = \frac{F_{b}}{F_{a}} \xi_{a}^{A,A'} + K_{b} - \frac{F_{b}}{F_{a}} K_{a},
\end{equation}
where $b$ represents a second transition (in the same atom).  One therefore expects a linear relationship between the isotope shifts of one transition against those of another, in the form of a King plot.
In the case of mercury,
the isotope shifts  between isotopes 198, 199, and 201 (from Table \ref{tab:corrected Hg line shifts}) 
are  used to construct a King plot with two data points. The isotope shifts are mass-scaled as in Table \ref{tab:Hg modified isotope shifts}, before creating the King plot 
between the clock and intercombination  lines  as seen in Fig.~\ref{fig:Hg King plot}.

\begin{table}[h]
    \centering
    \caption{Mass-scaled isotope shifts for the \clock\ clock and \ICL\ intercombination lines in Hg \textsc{i}. $\xi^{A,198} = \delta\nu^{A,198}/\mu^{A,198}$. 
    }
    \begin{ruledtabular}
    \begin{tabular}{ccd} 
        $A$ & Transition & \multicolumn{1}{c}{$\xi^{A,198}$\,(THz\,u)}\\ \hline
        199 & clock & 26.02(41)\\
        201 & clock & 84.64(14)\\
        196 & ICL & 81.375(78)\\
        199 & ICL & 25.65(25)\\
        200 & ICL & 95.0173(66)\\
        201 & ICL & 84.872(75)\\
        202 & ICL & 100.8337(28)\\
        204 & ICL & 102.8895(21)\\ 
    \end{tabular}
    \end{ruledtabular}
    \label{tab:Hg modified isotope shifts}
\end{table}

A linear fit to the King plot is calculated using an orthogonal distance regression \cite{Boggs1990} to account for errors in both ordinate and abscissa values. The gradient of the fit is found to be $1.0103(87)$ and the intercept to be $-0.64(70)\,\mathrm{THz\,u}$, where the uncertainties are taken to be the square roots of the diagonal entries in the covariance matrix of fit parameters
\begin{align}\label{eq:Hg covariance}
   & \Sigma_{\mathrm{fit}} =   \nonumber \\
   & \left(\begin{array}{cc}
        7.41160\times 10^{-5} & -5.85175\, \mathrm{GHz\,u} \\
        -5.85175\, \mathrm{GHz\,u}  &   0.484327\,  \mathrm{THz}^2 \mathrm{u}^{2}
    \end{array}\right).  
\end{align}
The unknown mass-scaled isotope shifts for clock transition are interpolated from Fig.~\ref{fig:Hg King plot} based on the gradient and intercept values, and appear in Table \ref{tab:Hg interpolated modified isotope shifts}. 
These are used in Sec.~\ref{sec:ClockFreqs} to predict the unexplored clock transition frequencies.  In estimating the uncertainties, the negative covariance between the gradient and the intercept of the linear fit is accounted for. 

\begin{table}[h]
    \centering
    \caption{Interpolated mass-scaled isotope shifts for the \clock\ clock line in Hg \textsc{i} assuming linearity in the King plot of Fig.~\ref{fig:Hg King plot}.}
      \begin{ruledtabular}
    \begin{tabular}{cd} 
        $A$ & \multicolumn{1}{c}{$\xi_{\mathrm{clock}}^{A,198}$\,(THz\,u)}\\ \colrule 
        196 & 81.18(17)\\
        200 & 94.68(20)\\
        202 & 100.44(24)\\
        204 & 102.47(25)\\  
    \end{tabular}
      \end{ruledtabular}
    \label{tab:Hg interpolated modified isotope shifts}
\end{table}

In the case of Cd there is insufficient data in the literature to produce an equivalent King plot for the clock and intercombination lines.  However, we can compute the slope and intercept values from our atomic structure calculations (following Section).

\section{\label{sec:IS}Isotope shift parameters}

Following the MCDHF-CI computation of the atomic state functions, the electronic mass shift and field shift parameters for the isotope shifts are calculated using \textsc{ris4} \cite{Ekman2019}. This is done for each stable isotope for both Hg and Cd.  A slight isotope dependence is observed for both elements. These values for Hg are presented in Table \ref{tab:Hg IS parameters} along with literature values for comparison. Our uncertainties for each isotope are prescribed using the  method 
outlined in Appendix \ref{app:Uncertainties}. The uncertainties for the mean values are calculated by adding the uncertainty for an individual isotope and the standard deviation of the set of isotopic values in quadrature.

\begin{table}[h]
    \centering
    \caption{Isotope shift parameters for the \clock\ clock and \ICL\ intercombination lines in Hg \textsc{i}. Repeated values amongst the isotopes are indicated by the ditto symbol $('')$.}
    \begin{threeparttable}
     \begin{ruledtabular}
    \begin{tabular}{ccccc} 
        $A$ & $F_{\mathrm{clock}}$ & $F_{\mathrm{ICL}}$ & $K_{\mathrm{clock}}$ & $K_{\mathrm{ICL}}$ \\ 
            $ $  & (GHz\,fm$^{-2}$) &  (GHz\,fm$^{-2}$)  & (THz\,u) & (THz\,u) \\\hline
        196 & -57.78(98) & -58.14(99) & -1.19(18) & -1.15(17)\\
        198 & -57.75(98) & -58.10(99) & \multicolumn{1}{c}{$''$} & \multicolumn{1}{c}{$''$}\\
        199 & -57.74(98) & -58.10(99) & \multicolumn{1}{c}{$''$} & \multicolumn{1}{c}{$''$}\\
        200 & -57.71(98) & -58.07(99) & \multicolumn{1}{c}{$''$} & \multicolumn{1}{c}{$''$}\\
        201 & -57.70(98) & -58.05(99) & \multicolumn{1}{c}{$''$} & \multicolumn{1}{c}{$''$}\\
        202 & -57.67(98) & -58.02(99) & \multicolumn{1}{c}{$''$} & \multicolumn{1}{c}{$''$}\\
        204 & -57.62(98) & -57.98(99) & \multicolumn{1}{c}{$''$} & \multicolumn{1}{c}{$''$}\\ \hline
        Mean & -57.71(99) & -58.07(1.00) & -1.19(18) & -1.15(17)\\
        Ref. \cite{Fricke2004_Hg} &  & -53.0 &  & 0.575\tnote{a}\\
        Ref. \cite{Angeli2013} &  & -55.36\tnote{b} &  & -0.65(33)\tnote{c}\\
        Ref. \cite{Ulm1986} &  & -55.4(3.9) &  & -0.6(0.3)\tnote{c}\\
        Ref. \cite{Lee1978} &  & -55.1(3.1) &  & \\
        Ref. \cite{Emrich1981} &  & -57.4(2.9) &  & \\ 
    \end{tabular}
     \end{ruledtabular}
    \begin{tablenotes}\footnotesize
    \item[a]{Normal mass shift factor only}
    \item[b]{Computational value from \cite{Torbohm1985}}
    \item[c]{Normal mass shift factor with specific mass shift of $(0\pm0.5)$. NMS; negative by the sign convention adopted here}
    \end{tablenotes}
    \end{threeparttable}
    \label{tab:Hg IS parameters}
\end{table}

The uncertainties calculated for Hg are found to be more than an order of magnitude larger than those for Yb. This is  not due a larger variation across the isotopes, but
more likely due to poorer convergence from either the inclusion of $5d$-orbitals or the SR  approach inflating the uncertainty for an individual isotope. Our value of $F_{\mathrm{ICL}}$ is the largest in magnitude of the reported values in the table; however, it is consistent with all of the values with published uncertainties to within 1\,$\sigma$. 
All the literature values of $K_{\mathrm{ICL}}$  have no contribution from the specific mass shift, whereas our value includes contributions from both the one-body (normal) and two-body (specific) mass shift terms. Ref. \cite{Emrich1981} finds the specific mass shift determined  from a King plot to be 14 times larger than the normal mass shift, and with an error of about 70\,\%, corresponding to a  large $K_\mathrm{ICL}$ value with large uncertainty and so this value has not been listed.

The gradient of the King plot in Figure \ref{fig:Hg King plot} is given by $F_{\mathrm{ICL}} / F_{\mathrm{clock}}$ and the intercept by $K_{\mathrm{ICL}} - K_{\mathrm{clock}} F_{\mathrm{ICL}} / F_{\mathrm{clock}}$. The calculated values in Table \ref{tab:Hg IS parameters} thus give a gradient of $1.006(25)$ and an intercept of $0.04(26)$\,THz\,u. These values are consistent with the fit parameters extracted from the King plot in Section \ref{sec:KingP} at the level of uncertainty.

The isotope shift parameters and their uncertainties are calculated for Cd by a similar means as for Hg. These values and their averages are presented in Table \ref{tab:Cd IS parameters} along with previously computed values for comparison. 
Field shift values calculated using a MR set for the zeroth order wave function were found to be inconsistent with the values from the SR  approach and from  literature values, hence why we adopted the SR method. 

\begin{table}[h]
    \centering
    \caption{Isotope shift parameters for the \clock\ clock and \ICL\ intercombination lines in Cd \textsc{i}.}
    \begin{threeparttable}
     \begin{ruledtabular}
    \begin{tabular}{ccccc} 
          $A$ & $F_{\mathrm{clock}}$ & $F_{\mathrm{ICL}}$ & $K_{\mathrm{clock}}$ & $K_{\mathrm{ICL}}$ \\ 
            $ $  & (GHz\,fm$^{-2}$) &  (GHz\,fm$^{-2}$)  & (THz\,u) & (THz\,u) \\\hline
        106 & -4.67(16) & -4.68(16) & -1.629(19) & -1.616(17)\\
        108 & \multicolumn{1}{c}{$''$} & \multicolumn{1}{c}{$''$} & \multicolumn{1}{c}{$''$} & \multicolumn{1}{c}{$''$}\\
        110 & -4.66(16) & \multicolumn{1}{c}{$''$} & \multicolumn{1}{c}{$''$} & \multicolumn{1}{c}{$''$}\\
        111 & \multicolumn{1}{c}{$''$} & \multicolumn{1}{c}{$''$} & \multicolumn{1}{c}{$''$} & \multicolumn{1}{c}{$''$}\\
        112 & \multicolumn{1}{c}{$''$} & \multicolumn{1}{c}{$''$} & \multicolumn{1}{c}{$''$} & \multicolumn{1}{c}{$''$}\\
        113 & \multicolumn{1}{c}{$''$} & -4.67(16) & \multicolumn{1}{c}{$''$} & \multicolumn{1}{c}{$''$}\\
        114 & \multicolumn{1}{c}{$''$} & \multicolumn{1}{c}{$''$} & \multicolumn{1}{c}{$''$} & \multicolumn{1}{c}{$''$}\\
        116 & \multicolumn{1}{c}{$''$} & \multicolumn{1}{c}{$''$} & \multicolumn{1}{c}{$''$} & \multicolumn{1}{c}{$''$}\\ \hline
        Average & -4.66(16) & -4.68(16) & -1.629(19) & -1.616(17)\\
        Ref. \cite{Fricke2004_Cd} &  & -4.42(34) &  & -1.72(33)\\
        Ref. \cite{Buchinger1987} &  & -3.91(46)\tnote{a} &  & -0.31(41)\tnote{a}\\
        Ref. \cite{Angeli2013} &  &  &  & -0.88(23)\tnote{a}\\  
        Ref. \cite{Torbohm1985} &  & -4.162 &  & \\
        Ref. \cite{Libert2007} &  & -4.37(18) &  & -1.72(18)\tnote{b}\\ 
    \end{tabular}
     \end{ruledtabular}
    \begin{tablenotes}\footnotesize
      \item[a]{The value is positive in the reference, but becomes negative  according to the isotope shift and mass shift conventions adopted here.}
    \item[b]{The value is presented as a ratio of specific mass shift to normal mass shift, hence leads to a negative total mass shift according to the convention adopted here.}
    \end{tablenotes}
    \end{threeparttable}
    \label{tab:Cd IS parameters}
\end{table}

The uncertainties calculated for Cd are again found to be larger  than for Yb.
This is also likely due to worse convergence of either having $4d$-orbitals just beneath the valence $5s$-orbitals or due to the SR set approach having larger variation during convergence for each isotope. Our value of $F_{\mathrm{ICL}}$ is  again the largest (in magnitude)  in comparison to previously estimated values, but is within two  standard deviations of all other values where uncertainties are presented.
The literature values of $K_{\mathrm{ICL}}$ display dichotomy between being consistent with our value (Refs.~\cite{Fricke2004_Cd,Libert2007}) and inconsistent with our value (Refs.~\cite{Buchinger1987,Angeli2013}).

In the case of Yb~\cite{Schelfhout2021(2)} and Hg, here, we have demonstrated that experimental data and atomic structure computations yield consistent values for the gradient and intercept in a King plot.  Therefore, despite the inability to construct a King plot between the clock and ICL transitions of Cd using experimental values, we can still infer the gradient and intercept values from our computed $K$ and $F$ values. The calculated values in Table \ref{tab:Cd IS parameters} give a gradient of $1.004(49)$ and an intercept of $0.02(9)$\,THz\,u for $\xi_\mathrm{ICL}^{A,A'}$ along the vertical axis. These are used later to predict clock line frequencies in Cd.

\section{\label{sec:Nuclear}Nuclear charge parameters}

The isotope shift equation (\ref{eq:isotope shift}) can be used to determine values for $\lambda^{A,A'}$ from measured ICL isotope shifts (Table \ref{tab:corrected Hg line shifts}), atomic masses (Table \ref{tab:Hg nuclear properties}), and calculated mass shift and field shift parameters (Table \ref{tab:Hg IS parameters}), where the tables here are specific to Hg. 
The resulting $\lambda^{A,A'}$ values for Hg are listed in Table \ref{tab:Hg nuclear parameters}, together with literature values for comparison. The uncertainties are calculated using propagation of errors under the assumption that the uncertainties in $F_{\mathrm{ICL}}$, $K_{\mathrm{ICL}}$, $\delta\nu_{\mathrm{ICL}}^{A,A'}$, and $\mu^{A,A'}$ are independent, which is likely an approximation since $F_{\mathrm{ICL}}$ and $K_{\mathrm{ICL}}$ are calculated from the same atomic state functions. Our values have the highest precision, but are mostly smaller (in magnitude) than the literature values due to our larger mass shift factor that includes the specific mass shift.  Our uncertainties are dominated by the uncertainty in $F_{\mathrm{ICL}}$.

\begin{table}[h]
    \centering
    \caption{Nuclear charge parameters {$\lambda^{A,198}$}  for Hg \textsc{i}.}
    \begin{ruledtabular}
    \begin{tabular}{cD{.}{.}{3.6}D{,}{}{3.4}D{.}{.}{3.6}D{,}{}{3.4}} 
         $A$& \multicolumn{4}{c}{$\lambda^{A,198}$\,($10^{-3}\,\mathrm{fm}^2$)}\\ 
         & \multicolumn{1}{c}{This work} & \multicolumn{1}{c}{Ref. \cite{Lee1978}} & \multicolumn{1}{c}{Ref. \cite{Fricke2004_Hg}} & \multicolumn{1}{c}{Ref. \cite{Bonn1976}} \\ \hline
        196 & -73.3\rlap{(1.3)} &  &  & -92,\rlap{(62)}\\
        198 & 0 & 0, & 0 & 0,\\
        199 & 11.72\rlap{(24)} & 27,\rlap{(12)} &  & 15,\rlap{(54)}\\
        200 & 83.8\rlap{(1.5)} & 103,\rlap{(26)} & 91.8\rlap{(4.0)} & 108,\rlap{(47)} \\
        201 & 111.9\rlap{(2.0)} & 116,\rlap{(17)} &  & 144,\rlap{(44)}\\
        202 & 175.9\rlap{(3.1)} & 199,\rlap{(25)} & 195.3\rlap{(5.7)} & 226,\rlap{(41)}\\
        204 & 266.6\rlap{(4.7)} & 298,\rlap{(28)} & 292.7\rlap{(7.0)} & 343,\rlap{(38)}\\ 
    \end{tabular}
    \end{ruledtabular}
    \label{tab:Hg nuclear parameters}
\end{table}

\begin{figure}[h]
    \centering
    \includegraphics[width=1.02\columnwidth]{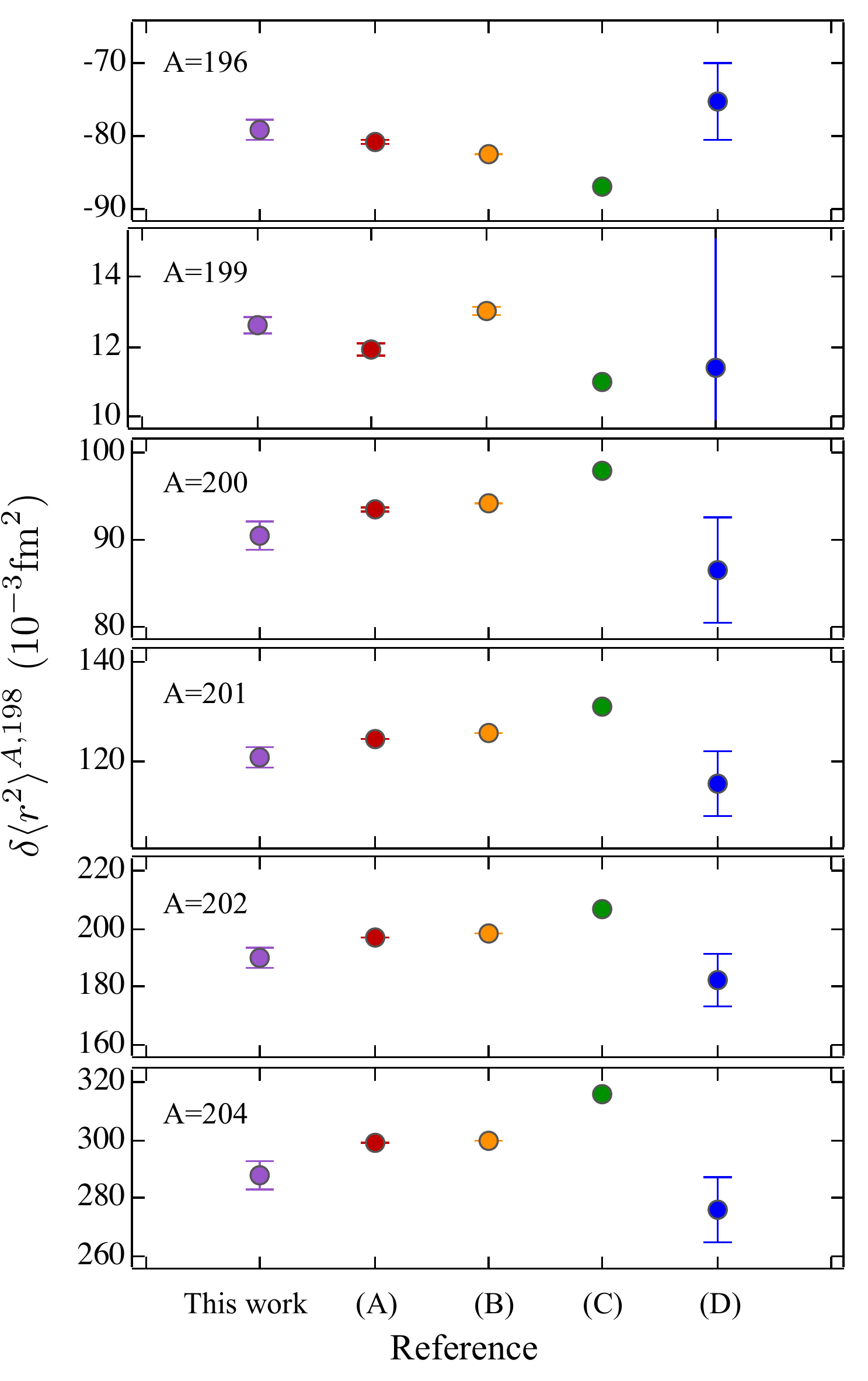} 
    \caption{Differences in mean-square nuclear charge radii $\delta\langle r^2\rangle^{A,198}$ for isotopes in Hg.  The references are as follows: (A) Ulm~\ea~Ref.~\cite{Ulm1986}, (B) Angeli and Marinova~\cite{Angeli2013}, (C) Fricke and Heilig~\cite{Fricke2004_Hg}, and (D) Rayman~\ea~\cite{Rayman1989}. 
    }
    \label{fig:Hg nuclear plots}
\end{figure}

Values for the nuclear charge parameter $\lambda^{A,A'}$ can be converted into differences in mean-square nuclear charge radii $\delta\langle r^2\rangle^{A,A'}$ through the use of a scaling factor to account for the higher-order terms \cite{Schelfhout2021(2),Fricke2004_Hg}. For Hg this relationship is taken to be $\delta\langle r^2\rangle^{A,A'} = \lambda^{A,A'}/0.927$ \cite{Fricke2004_Hg} by evaluating the higher-order nuclear moments from electron scattering data and quantifying their contribution with Seltzer's coefficients. The calculated differences in mean-square nuclear charge radii are presented in Table \ref{tab:Hg delta r squared} along with literature values, and are  plotted in Fig.~\ref{fig:Hg nuclear plots} for a visual comparison. Our values lie between other values from the literature and are consistent with the experimental values of Rayman \textit{et al.} \cite{Rayman1989} within uncertainties. The  dominant sources of uncertainty for $\delta\langle r^2\rangle^{A,A'}$ estimates are associated with the mass- and field-shift parameters, which appear to be neglected in the values from Refs.~\cite{Ulm1986,Angeli2013}. 
Precise values for the differences in mean-square nuclear charge radii are important in constraining models used in nuclear structure calculations \cite{Reinhard2021,Ma2020,Reinhard2020,Papoulia2016}. The consistency of our values with  
prior values, and our higher precision (where uncertainties for  $K$ and $F$ are included) suggests that our values should be useful for constraining nuclear models  relevant to computations involving the stable isotopes of mercury.  

\begin{table}[h]
    \centering
    \caption{Differences in mean-square nuclear charge radii {$\delta\langle r^2\rangle^{A,198}$} determined by $\delta\langle r^2\rangle^{A,A'} = \lambda^{A,A'}/0.927$ for Hg \textsc{i} .} 
    \begin{threeparttable}
    \begin{ruledtabular}
    \begin{tabular}{cD{.}{.}{3.6}D{.}{.}{3.6}D{.}{.}{3.6}D{,}{}{3.0}D{.}{.}{3.7}} 
       $A$ & \multicolumn{5}{c}{$\delta\langle r^2\rangle^{A,198}$\,($10^{-3}\,\mathrm{fm}^2$)}\\ 
        & \multicolumn{1}{c}{This work} & \multicolumn{1}{c}{Ref. \cite{Ulm1986}\tnote{a}} & \multicolumn{1}{c}{Ref. \cite{Angeli2013}\tnote{a}} & \multicolumn{1}{c}{Ref. \cite{Fricke2004_Hg}} & \multicolumn{1}{c}{Ref. \cite{Rayman1989}}\\ \hline
        196 & -79.1\rlap{(1.4)} & -80.9\rlap{(0.3)} & -82.5\rlap{(0.1)} & -87, & -75.2\rlap{(5.4)}\\
        198 & 0 & 0 & 0 & 0, & 0\\
        199 & 12.64\rlap{(26)} & 11.9\rlap{(0.2)} & 13.0\rlap{(0.1)} & 11, & 11.4\rlap{(8.1)}\\
        200 & 90.4\rlap{(1.6)} & 93.5\rlap{(0.2)} & 94.2\rlap{(0.1)} & 98, & 86.6\rlap{(6.1)}\\
        201 & 120.7\rlap{(2.1)} & 124.5\rlap{(0.2)} & 125.8\rlap{(0.1)} & 131, & 115.6\rlap{(6.5)}\\
        202 & 189.8\rlap{(3.3)} & 197.0\rlap{(0.2)} & 198.1\rlap{(0.1)} & 207, & 182.1\rlap{(9.2)}\\
        204 & 287.6\rlap{(5.0)} & 298.8\rlap{(0.3)} & 300.1\rlap{(0.1)} & 316, & 276.1\rlap{(11.4)}\\ 
    \end{tabular}
    \end{ruledtabular}
    \begin{tablenotes}\footnotesize
    \item[a]{Uncertainties represent experimental errors only and neglect the additional uncertainties from $K$ and $F$}
    \end{tablenotes}
    \end{threeparttable}
    \label{tab:Hg delta r squared}
\end{table}

Using a similar procedure as for Hg, values for $\lambda^{A,A'}$ for Cd can be determined using the values from Tables \ref{tab:corrected Cd line shifts}, \ref{tab:Cd nuclear properties} and \ref{tab:Cd IS parameters}. The resulting nuclear charge parameters are presented in Table \ref{tab:Cd nuclear parameters} along with literature values. The uncertainties are calculated using propagation of errors assuming independent errors. Major error contributors include the uncertainty in $F_{\mathrm{ICL}}$ (again), and also  the uncertainty of  the measured isotopes shifts. The relative strengths of these errors depends on the isotope. Our values lie between the values of Refs.~\cite{Fricke2004_Cd} (Fricke and Heilig) and \cite{Buchinger1987} (Buchinger~\ea), and are consistent with them to within $2\sigma$ and $1\sigma$,  respectively (where $1\sigma$ is one standard deviation). 
The high degree of consistency between our values and those calculated from Ref.~\cite{Angeli2013} is to be expected, since 
the experimental isotope shifts from \cite{Fricke2004_Cd} are used to calculate $\lambda^{A,A'}$ in this work and in Ref.~\cite{Angeli2013}. 

\begin{table}[h]
    \centering
    \caption{Nuclear charge parameters {$\lambda^{A,114}$} for Cd \textsc{i}.}
    \begin{threeparttable}
    \begin{ruledtabular}
    \begin{tabular}{cD{.}{.}{2.8}D{.}{.}{2.8}D{.}{.}{2.7}D{.}{.}{2.7}} 
      $A$ & \multicolumn{4}{c}{$\lambda^{A,114}$\,($\mathrm{fm}^2$)}\\
        & \multicolumn{1}{c}{This work} & \multicolumn{1}{c}{Ref. \cite{Fricke2004_Cd}} & \multicolumn{1}{c}{Ref. \cite{Angeli2013}\tnote{b}} & \multicolumn{1}{c}{Ref. \cite{Buchinger1987}\tnote{c}}\\ \hline
        106 & -0.645(23) & -0.6712(84) & -0.646(86) & -0.535(97)\\
        108 & -0.465(19) & -0.4988(73) & -0.464(63) & -0.392(71)\\
        110 & -0.301(12) & -0.3244(60) & -0.300(41) & -0.256(47)\\
        111 & -0.261(14) & -0.2807(73) & -0.268(37) & -0.235(41)\\
        112 & -0.1408(55) & -0.1549(42) & -0.139(20) & -0.121(23)\\
        113 & -0.0981(94) & -0.1102(60) & -0.103(17) & -0.099(17)\\
        114 & 0 & 0 & 0 & 0\\
        116 & 0.1121(47) & 0.1294(42) & 0.105(16) & 0.094(20)\\ 
    \end{tabular}
    \end{ruledtabular}
    \begin{tablenotes}\footnotesize
    \item[b]{Values calculated as per data and method in Ref. \cite{Angeli2013}; uncertainties from $K$ and $F$ incorporated}
    \item[c]{Presented as $\delta\langle r^2\rangle$, but higher-order terms have been ignored in their determination, so these values are effectively $\lambda^{A,114}$}
    \end{tablenotes}
    \end{threeparttable}
    \label{tab:Cd nuclear parameters}
\end{table}

\begin{table*}[htbp]  
    \centering
    \caption{Differences in mean-square nuclear charge radii {$\delta\langle r^2\rangle^{A,114}$} for Cd determined by $\delta\langle r^2\rangle^{A,A'} = \lambda^{A,A'}/0.973$.} 
    \begin{threeparttable}
    \begin{ruledtabular}
    \begin{tabular}{cD{.}{.}{2.8}D{.}{.}{2.7}D{.}{.}{2.7}D{.}{.}{2.7}D{.}{.}{2.8}D{.}{.}{2.7}D{.}{.}{2.3}} 
        $A$& \multicolumn{6}{c}{$\delta\langle r^2\rangle^{A,114}$\,($\mathrm{fm}^2$)}\\ 
        1 &\multicolumn{1}{c}{2} & \multicolumn{1}{c}{3} & \multicolumn{1}{c}{4} & \multicolumn{1}{c}{5} & \multicolumn{1}{c}{6} & \multicolumn{1}{c}{7} & \multicolumn{1}{c}{8} \\  
        & \multicolumn{1}{c}{This work} & \multicolumn{1}{c}{Ref. \cite{Angeli2013}\tnote{a}} & \multicolumn{1}{c}{Ref. \cite{Angeli2013}\tnote{b}} & \multicolumn{1}{c}{Ref. \cite{Fricke2004_Cd}} & \multicolumn{1}{c}{Ref. \cite{Buchinger1987}\tnote{c}}& \multicolumn{1}{c}{Ref. \cite{Hammen2018}} & \multicolumn{1}{c}{Ref. \cite{Libert2007}\tnote{d}}\\ \hline
        106 & -0.662(24) & -0.576(8) & -0.663(89) & -0.696(28) & -0.550(100)& -0.695(13) & -0.644\\
        108 & -0.478(19) & -0.412(11) & -0.477(65) & -0.514(23) & -0.403(73) & -0.510(10) & -0.477\\
        110 & -0.310(12) & -0.252(5) & -0.308(42) & -0.331(17) & -0.263(49) & -0.334(7) & -0.304\\
        111 & -0.269(15) & -0.130(13) & -0.275(38) & -0.285(21) & -0.242(43) & -0.288(13) & \\
        112 & -0.1448(56) & -0.103(3) & -0.143(20) & -0.157(10) & -0.124(24) & -0.159(5) & -0.126\\
        113 & -0.1008(97) & -0.008(4) & -0.105(18) & -0.111(17) & -0.102(18) & -0.114(11) & \\
        114 & 0 & 0 & 0 & 0 & 0 & 0 & 0\\
        116 & 0.1152(48) & 0.088(3) & 0.108(17) & 0.129(10) & 0.097(21) & 0.134(9) & 0.109\\ 
    \end{tabular}
    \end{ruledtabular}
    \begin{tablenotes}\footnotesize
    \item[a]{Uncertainties represent statistical errors only and neglect the uncertainties for $K$ and $F$.}
    \item[b]{Values re-calculated (with higher-order terms as in this work) using experimental isotope shifts and $K$ and $F$ values as in Ref. \cite{Angeli2013}; uncertainties from $K$ and $F$ included.}
    \item[c]{Higher-order terms have been accounted for as per the method in \cite{Fricke2004_Cd}.}
    \item[d]{Values arising from nuclear structure calculations.}
    \end{tablenotes}
    \end{threeparttable}
    \label{tab:Cd delta r squared}
\end{table*}

\begin{figure}[htbp]
    \centering
    \includegraphics[width=1.02\columnwidth]{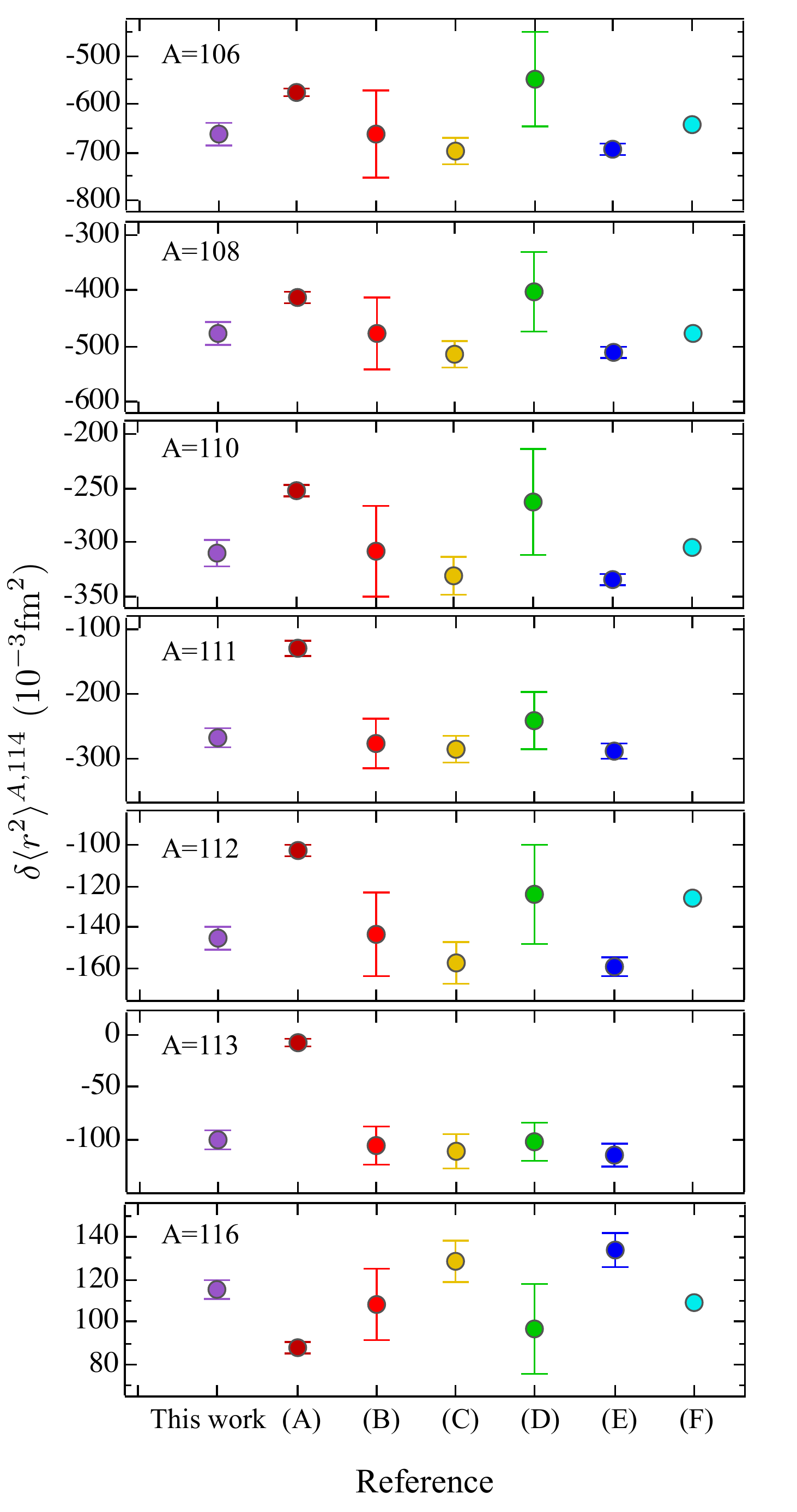} 
    \caption{Differences in  mean-square nuclear charge radii $\delta\langle r^2\rangle^{A,114}$ for  isotopes in Cd from this work and previous sources. The references are as follows: (A) Angeli and Marinova~\cite{Angeli2013}, (B) Ref. \cite{Angeli2013} but with recalculated values using the same inputs as Angeli and Marinova, (C) Fricke and Heilig~\cite{Fricke2004_Cd}, (D) Buchinger~\ea~\cite{Buchinger1987}, (E) Hammen~\ea~\cite{Hammen2018}, and (F) Libert~\ea~\cite{Libert2007}.
    }
    \label{fig:Cd_nuclear_plots}
\end{figure}

For Cd, the nuclear charge parameters can be converted into differences in mean-square nuclear charge radii as $\delta\langle r^2\rangle^{A,A'} = \lambda^{A,A'}/0.973$ \cite{Fricke2004_Cd}, where again the scaling factor is found by evaluating the higher-order nuclear moments from electron scattering data and quantifying their contribution with Seltzer's coefficients. The calculated differences in mean-square nuclear charge radii, with respect to $^{114}$Cd,  are presented in Table \ref{tab:Cd delta r squared} along with some literature values, and plotted in Fig.~\ref{fig:Cd_nuclear_plots}. Columns 3 to 8 of Table \ref{tab:Cd delta r squared} correspond to reference labels (A) to (F) in Fig.~~\ref{fig:Cd_nuclear_plots}. Our values are consistent with those of Refs.~\cite{Fricke2004_Cd,Buchinger1987,Libert2007} within uncertainties other than $\delta\langle r^2\rangle^{112,114}$ from \cite{Libert2007}. But the general agreement with Ref. \cite{Libert2007} is promising, since their values arise from nuclear structure calculations.
We note the values of $\delta\langle r^2\rangle^{111,114}$  and $\delta\langle r^2\rangle^{113,114}$ from the Data Tables of Angeli and Marinova~\cite{Angeli2013} lie several standard deviations away from the other values (seen clearly in Fig.~\ref{fig:Cd_nuclear_plots}). 
Column 4 of Table~\ref{tab:Cd delta r squared}  and label (B) in Fig. \ref{fig:Cd_nuclear_plots} are values recomputed based on  
the experimental isotope shifts, and $K$ and $F$ parameters that the authors (apparently) used for the preceding column. There is a difference in the method used to convert $\lambda^{A,A'}$ into $\delta\langle r^2\rangle^{A,A'}$: Ref. \cite{Angeli2013} use an iterative procedure described in Ref.~\cite{Angeli2004}, whereas we applied a re-scaling factor as in Ref.~\cite{Fricke2004_Cd}. We have propagated the uncertainties in the mass shift and field shift factors to arrive at the uncertainties in these corrected values. These recomputed values  are in  good agreement with our values. 

The most precise values for $\delta\langle r^2\rangle^{A,A'}$  are listed in column 7 from Hammen~\ea~\cite{Hammen2018}.  These are in close agreement with Fricke and Heilig~\cite{Fricke2004_Cd} (col. 5), but are slightly inconsistent with our values and the nuclear structure calculations of Libert~\ea~\cite{Libert2007}.  The Hammen~\ea\ values rely on frequency measurements of the $5s5p\,\,^3P_2 - 5s6s\,\,^3S_1$ transition from Ref.~\cite{Frommgen2015}, and obtain their own $K$ and $F$ values through a King plot procedure involving Cd~\textsc{ii}. 

\section{\label{sec:ClockFreqs}Clock transition frequencies}

Our previous work with Yb presented clock transition isotope shifts determined using three largely independent methods \cite{Schelfhout2021(2)}. The method of scaling the \textit{ab initio} isotope shifts (in Table \ref{tab:Hg clock ab initio energies}) to match the experimental isotope shifts (of Table \ref{tab:corrected Hg line shifts}) is not performed here as it is considered the least reliable method and, unlike for Yb, there does not appear to be a common factor that appropriately scales the \textit{ab initio} isotope shifts to agree with the experimental values. Isotope shifts, $\delta\nu_\mathrm{clock}^{A,A'}$, for the clock transition are calculated from the isotope shifts of the ICL in Table \ref{tab:corrected Hg line shifts} using the two more reliable methods --- using the computational values from Table \ref{tab:Hg IS parameters} in Eq.~(\ref{eq:King}) and undoing the mass-scaling of the interpolated values from Fig.~\ref{fig:Hg King plot} in Table \ref{tab:Hg interpolated modified isotope shifts}. These isotope shifts are presented in Table \ref{tab:Hg clock isotope shifts}.

\begin{table}[h]
    \centering
    \caption{Isotope shifts $\delta\nu_{\mathrm{clock}}^{A,198}$  for the \clock\ clock line in Hg \textsc{i}. Column 2 gives values calculated using the isotope shift parameters in Table \ref{tab:Hg IS parameters}; column 3 gives values calculated using the interpolations from the King plot in Table \ref{tab:Hg interpolated modified isotope shifts} and the known values for ${}^{199,201}$Hg are also re-calculated for consistency.}
       \begin{ruledtabular}
    \begin{tabular}{cD{.}{.}{3.7}D{.}{.}{3.8}} 
       $A$ & \multicolumn{2}{c}{$\delta\nu_{\mathrm{clock}}^{A,198}$\,(GHz)}\\ 
        & \multicolumn{1}{c}{Atomic structure calculations} & \multicolumn{1}{c}{King plot linearity}\\ \hline
        196 & 4.17(11) & 4.1888(87)\\
        198 & 0 & 0\\
        199 & -0.646(19) & -0.662(14)\\
        200 & -4.77(12) & -4.789(11)\\
        201 & -6.37(16) & -6.392(14)\\
        202 & -10.03(25) & -10.062(24)\\
        204 & -15.21(38) & -15.250(38)\\ 
    \end{tabular}
    \end{ruledtabular}
    \label{tab:Hg clock isotope shifts}
\end{table}

The uncertainties in the isotope shifts are calculated by propagation of uncertainties, where the covariance matrix (\ref{eq:Hg covariance}) is used for the third column. 
The large negative covariance between the gradient and intercept of the fit lead to the uncertainties from the King plot interpolation being an order of magnitude smaller than for the method involving atomic structure calculations (for which covariance is ignored). An inverse-variance weighted average of the values in Table~\ref{tab:Hg clock isotope shifts} is summed with  the $^{198}$Hg  absolute frequency from Table~\ref{tab:Hg absolute values} to arrive at the  clock transition frequencies for all the isotopes. These predicted values
are presented along with the literature values in the lower half of Table \ref{tab:predictions}.  The much more accurately measured value for $^{199}$Hg may seem a more suitable choice to sum the isotope shifts to, but this contains the second order hyperfine shift. 
The uncertainties have contributions from both the clock line frequency of ${}^{198}$Hg  and the isotope shifts, though in the case of ${}^{196}$Hg and ${}^{200}$Hg it is dominated by the former.

\begin{table}[h]
    \centering
    \caption{Predictions for the absolute transition frequencies of the \clock\ clock line in isotopes of Cd \textsc{i} and Hg \textsc{i}.}
    \begin{ruledtabular}
    \begin{tabular}{ldc} 
        Isotope & \multicolumn{1}{c}{$\nu_{\mathrm{clock}}$\,(GHz)} & Reference\\ \hline
         ${}^{106}\mathrm{Cd}$ & 902796.54(17) & \\
        ${}^{108}\mathrm{Cd}$ & 902795.99(13) & \\
        ${}^{110}\mathrm{Cd}$ & 902795.503(93) & \\
        ${}^{111}\mathrm{Cd}$ & 902795.450(98) & \\
        ${}^{112}\mathrm{Cd}$ & 902795.019(69) & \\
        ${}^{113}\mathrm{Cd}$ & 902794.949(76) & \\
        ${}^{114}\mathrm{Cd}$ & 902794.618(60) & \cite{Burns1956}\\
        ${}^{116}\mathrm{Cd}$ & 902794.343(67) & \\ \hline
        ${}^{196}\mathrm{Hg}$ & 1128580.144(14) & \\
        ${}^{198}\mathrm{Hg}$ & 1128575.955(11) & \cite{Kramida2011}\\
        ${}^{199}\mathrm{Hg}$ & 1128575.29080815462(41) & \cite{Tyumenev2016}\\
        ${}^{200}\mathrm{Hg}$ & 1128571.167(16) & \\
        ${}^{201}\mathrm{Hg}$ & 1128569.5611396(53) & \cite{Petersen2008}\\
        ${}^{202}\mathrm{Hg}$ & 1128565.894(27) & \\
        ${}^{204}\mathrm{Hg}$ & 1128560.706(40) & \\ 
    \end{tabular}
    \end{ruledtabular}
    \label{tab:predictions}
\end{table}

With regard to Cd, in the absence of sufficient experimental data to construct a King plot, the most reliable method considered for calculating isotope shifts for the clock transition is the computational method detailed in Ref.~\cite{Schelfhout2021(2)}. The $K$ and $F$ values from Table \ref{tab:Cd IS parameters} are used in Eq.~(\ref{eq:King}) together with the ICL isotope shifts from Table~\ref{tab:corrected Cd line shifts} and the mass scaling is undone to arrive at the clock line isotope shifts presented in Table~\ref{tab:Cd clock isotope shifts}.
\begin{table}[h]
    \centering
    \caption{Isotope shifts for the \clock\ clock line in Cd \textsc{i}. 
    }
    \begin{ruledtabular}
    \begin{tabular}{cD{.}{.}{2.7}} 
        $A$ & \multicolumn{1}{c}{$\delta\nu_{\mathrm{clock}}^{A,114}$\, (GHz)} \\ \hline
        106 & 1.92(15) \\
        108 & 1.37(12) \\
        110 & 0.884(71) \\
        111 & 0.832(77) \\
        112 & 0.401(34) \\
        113 & 0.331(47) \\
        114 & 0 \\
        116 & -0.275(29)\\ 
    \end{tabular}
  \end{ruledtabular}
    \label{tab:Cd clock isotope shifts}
\end{table}

These isotope shifts are used to find the absolute clock transition frequencies given the  clock transition frequency of ${}^{114}$Cd (from Sec.~\ref{sec:compmethod}).   The predicted frequencies are listed with those of Hg in Table~\ref{tab:predictions}.
We note that the predictions for ${}^{112}$Cd, ${}^{113}$Cd, and ${}^{116}$Cd are limited by the uncertainty from ${}^{114}$Cd in Ref.~\cite{Burns1956} (and not the uncertainties on the $K$ and $F$ parameters computed here). %

In addition to the bosonic isotopes, we note that the fermionic isotopes also lack published experimental values for the clock transition frequency.  However, we have confirmed sub-$1\sigma$  agreement with experimental values that are accurate to at least  parts in $10^{10}$ for ${}^{111}\mathrm{Cd}$ and ${}^{113}\mathrm{Cd}$ through private communication with H.~Katori and A.~Yamaguchi from RIKEN, Tokyo.  

\section{\label{sec:conclusion}CONCLUSIONS}

\MCDHF\ calculations with \CI\ have been carried out for Hg \textsc{i} and Cd \textsc{i} with the \textsc{grasp2018} package. The resultant atomic wave functions have been used to determine the diagonal and off-diagonal hyperfine coupling constants in the fermionic isotopes with the \textsc{rhfs} program under \textsc{grasp2018}. For Hg and Cd, the diagonal hyperfine coupling constants  $A({}^3P_1^o)$  are  within 6\% and 0.04\% of values extracted from experimental data, respectively. 
We note, though, that the experimental values for Cd had 0.8\% uncertainty.
 For Hg, the \textit{ab initio} clock transition frequencies are at an error of only $0.02\%$ with experimental values, which appears to set a precedent for atomic structure calculations in Hg.  In the case of $^{114}$Cd there is a 0.8\% difference, which still represents a high level of agreement (the measured value was inferred from the difference between \ICL\ and ${}^3P_0^o - {}^3P_1^o$ frequencies).  
 
 The resultant wave functions from \textsc{grasp2018} are also used in the \textsc{ris4} routine to find mass ($K$) and field ($F$) shift parameters associated with isotope shifts.  We have evaluated $K$ and $F$ values for both the clock and intercombination lines in the respective elements, and by isotope.  This appears to be the first report of $K_\mathrm{clock}$ and $F_\mathrm{clock}$ for both Hg and Cd.  Our value for $F_\mathrm{ICL}$ for Hg is  larger in magnitude than previous reports with reduced uncertainty, but is consistent to within 1-$\sigma$.   Our Hg value for $K_\mathrm{ICL}$ includes contributions from normal and specific mass shifts and is approximately a factor of 2 greater than previously reported values, which only appear to include the normal mass shift.  For Cd, we find $F_\mathrm{ICL}$ to be slightly larger in magnitude than previous reports but agreeing to within uncertainties.  Our $K_\mathrm{ICL}$ value  agrees with some prior values but disagrees with others. 
 
 Calculated differences in mean-square nuclear charge radii are found to be consistent with previous values to within uncertainties for Hg. A King plot is constructed using available clock and intercombination transition data, and interpolating this yields isotope shifts for the clock transitions.  These isotope shifts are consistent with values deduced from mass- and field-shift parameters that are found via our atomic structure calculations.  The latter are less accurate, but the comparison gives confidence in the calculations.

 For Cd, the \textit{ab initio} clock transition frequencies are believed to be the only available calculations of this quantity. Calculated differences in mean-square nuclear charge radii are found to be mostly consistent with previously reported values. 
 There may be a case for the revision of the tabulated  $\delta\langle r^2\rangle^{A,A'}$ values for Cd in Ref.~\cite{Angeli2013}.  Notably, our values are, in the majority, consistent with nuclear structure calculations of Libert~\ea~\cite{Libert2007}.
 Predictions for the absolute clock transition frequencies for the fermionic isotopes of Cd are found to be consistent to below the 1-$\sigma$ range with experimental values communicated to us after these calculations were performed~\cite{Yamaguchi(private)}.

\begin{acknowledgments}
We  are indebted to  Christopher Bording and Hayden Walker from the UWA High Performance Computing Team. J.~S. acknowledges support from the John and Patricia Farrant Scholarship. This research was undertaken with the assistance of resources from the University of Western Australia High Performance Computing Hub.  We thank A. Yamaguchi and H. Katori for sharing with us preliminary measurements of the clock transition frequencies in $^{111}$Cd and $^{113}$Cd.  Thank you to Daniel Jones for the careful  proofreading of this manuscript.  
\end{acknowledgments}

\appendix

\section{\label{app:a} NUCLEAR PROPERTIES }

Data relevant to our calculations are listed in  Tables  \ref{tab:Hg nuclear properties} and  \ref{tab:Cd nuclear properties} for Hg and Cd, respectively.

\begin{table}[htbp]
    \centering
    \caption{Properties of Hg isotopes. $A$, mass number; $I$, nuclear spin; $\mu$, nuclear magnetic moment;  $R$, rms nuclear charge radius; $m_A$, neutral atomic mass. $\mu$ values are from \cite{Stone2005},  $R$ values are from Ref.~\cite{Angeli2013}, and $m_A$ values are from Ref.~\cite{Wang2021}. The only nonzero nuclear quadrupole moment occurs for ${}^{201}\mathrm{Hg}$, where $Q_I(b)= 0.387(6)$~\cite{Bieron2005}. For the abundances see Ref.~\cite{Meija2016}.
    }
    \begin{ruledtabular} 
    \begin{tabular}{ccD{.}{.}{2.12}D{.}{.}{1.8}D{.}{.}{3.10}}  
        $A$ & $I$ & \multicolumn{1}{c}{$\mu$ ($\mu_N$)} & \multicolumn{1}{c}{$R$ (fm)} & \multicolumn{1}{c}{$m_A$ (u)} \\ 
        \hline
        196 & 0 & 0  & 5.4385(31) & 195.965833(3) \\ 
        198 & 0 & 0  & 5.4463(31) & 197.9667692(5)\\ 
        199 & 1/2 & 0.5058855(9)  & 5.4474(31) & 198.9682810(6) \\ 
        200 & 0 & 0  & 5.4551(31) & 199.9683269(6) \\ 
        201 & 3/2 & -0.5602257(14) & 5.4581(32)   & 200.9703031(8) \\ 
        202 & 0 & 0  & 5.4648(33) & 201.9706436(8)\\ 
        204 & 0 & 0  & 5.4744(36) & 203.9734940(5) \\ 
    \end{tabular}
    \end{ruledtabular}  
    \label{tab:Hg nuclear properties}
\end{table}

\begin{table}[htbp]
    \centering
    \caption{Properties of Cd isotopes. $A$, mass number; $I$, nuclear spin; $\mu$, nuclear magnetic moment; $R$, rms nuclear charge radius; $m_A$, neutral atomic mass. $\mu$ values are from \cite{Stone2005}, $R$ values are from Ref. \cite{Angeli2013}, and $m_A$ values are from Ref. \cite{Wang2021}. For the abundances see Ref. \cite{Meija2016}.
    }
     \begin{ruledtabular}
    \begin{tabular}{ccD{.}{.}{2.10}D{.}{.}{1.8}D{.}{.}{3.12}D{.}{.}{2.7}} 
        $A$ & $I$ & \multicolumn{1}{c}{$\mu$ ($\mu_N$)} & \multicolumn{1}{c}{$R$ (fm)} & \multicolumn{1}{c}{$m_A$ (u)}\\ 
        \hline
        106 & 0 & 0 & 4.5383(36) & 105.9064598(12)\\ 
        108 & 0 & 0 & 4.5577(31) & 107.9041836(12)\\ 
        110 & 0 & 0 & 4.5765(26) & 109.9030075(4) \\ 
        111 & 1/2 & -0.5948861(8) & 4.5845(28) & 110.9041838(4) \\ 
        112 & 0 & 0 & 4.5944(24) & 111.90276390(27) \\ 
        113 & 1/2 & -0.6223009(9) & 4.6012(28) & 112.90440811(26) \\ 
        114 & 0 & 0 & 4.6087(23) & 113.90336500(30) \\ 
        116 & 0 & 0 & 4.6203(59) & 115.90476323(17) \\ 
    \end{tabular}
     \end{ruledtabular}
    \label{tab:Cd nuclear properties}
\end{table}

\section{\label{app:centroid} CENTROID CORRECTIONS }
For an isotope with nuclear spin $I=1/2$, the $J=1$ hyperfine manifold consists of only two levels: $F=1/2$ or $3/2$. To first-order, the shifts of these hyperfine levels are given by
\begin{align*}
    \delta\nu_F &= \frac{A}{2}(F(F+1)-I(I+1)-J(J+1)) \\
   &= \frac{A}{8}(4F(F+1)-11),
\end{align*}
where $A$ is the (diagonal) hyperfine interaction constant (there is no electric quadrupole term in this instance). For $F=1/2$ we have $\delta\nu = -A$ and for $F=3/2$, $\delta\nu = A/2$. These shifts are relative to the center of gravity of the hyperfine manifold, and so the   line frequencies are 
\begin{align*}
    \nu_{3/2} &= \nu_{\mathrm{c.o.g.}} + \frac{A}{2}\\
    \nu_{1/2} &= \nu_{\mathrm{c.o.g.}} - A.
\end{align*}
and the separation between hyperfine levels is $\Delta \nu_\mathrm{hfs}= \nu_{3/2}- \nu_{1/2}$.
Rearranging the pair of equations gives,
\begin{align*}
    \nu_{\mathrm{c.o.g.}} &= \frac{1}{3}(\nu_{1/2} + 2\nu_{3/2})\\
    A &= \frac{2}{3}(\nu_{3/2} - \nu_{1/2}).
\end{align*}

\begin{figure}[h]
    \centering  
    \includegraphics[width=1.0\columnwidth]{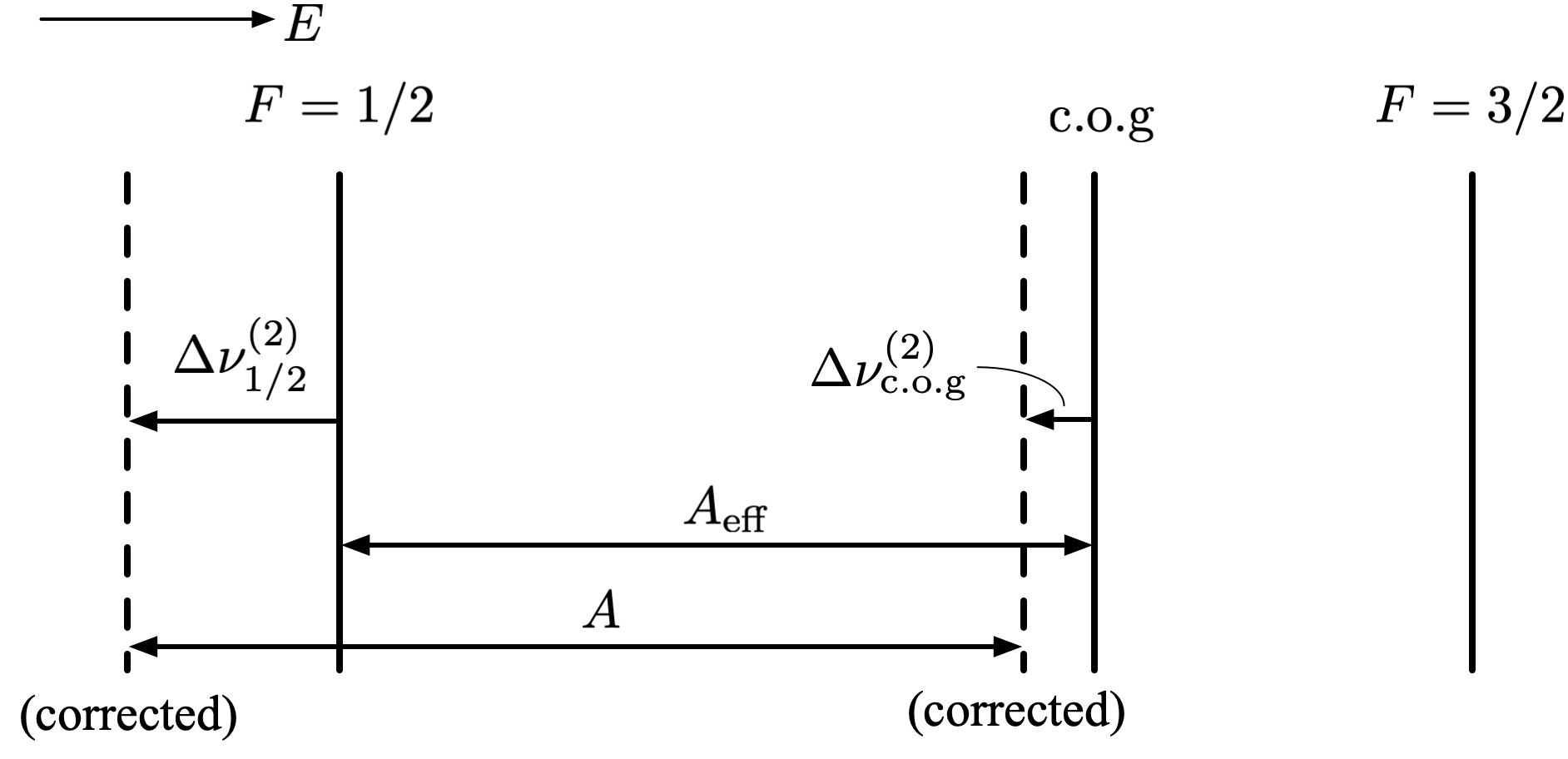} 
    \caption{Energy level structure of hyperfine manifold for $I=1/2, J=1$ including adjustments made to $F=1/2$ level due to the off-diagonal hyperfine interaction with the $J=0$ level. The $F=3/2$ level remains unperturbed. Solid lines represent measured values (and therefore uncorrected for a centroid comparison with bosonic isotope shifts). Positive energy to the right corresponds to positive $A$, as in the case of $^{199}$Hg. 
    }
    \label{fig:hyperfine}
\end{figure}

For $J=1$, the shift to the level $F=1/2$ due to the off-diagonal  hyperfine interaction with the $J=0$ level is positive, i.e. the measured frequency is higher than in the absence of the second-order hyperfine interaction. Adjusting for this perturbation is done by subtracting the second order shift $\Delta \nu_F^{(2)}$. This leads to the shifted value $\nu'_{1/2} = \nu_{1/2} - \Delta\nu_{1/2}^{(2)}$ and a shifted center of gravity
\begin{align*}
    \nu'_{\mathrm{c.o.g.}} &= \frac{1}{3}(\nu'_{1/2} + 2\nu_{3/2})\\
    &= \frac{1}{3}(\nu_{1/2} - \Delta\nu_{1/2}^{(2)} + 2\nu_{3/2})\\
    &= \nu_{\mathrm{c.o.g.}} - \frac{1}{3}\Delta\nu_{1/2}^{(2)}.
\end{align*}
And so the shift to the center of gravity $\Delta\nu_{\mathrm{c.o.g.}}^{(2)} = \nu_{\mathrm{c.o.g.}} - \nu'_{\mathrm{c.o.g.}}$ is equal to one-third of the $\Delta\nu_{F}^{(2)}$ shift of the $F=1/2$ level. This result is used in Sec.~\ref{sec:Centroid}. The energy levels are illustrated in Figure \ref{fig:hyperfine}. 

The measured hyperfine structure interval leads to an effective hyperfine interaction constant $A_{\mathrm{eff}}$ under the first-order theory. The diagonal hyperfine interaction constant, $A$, is recovered  by accounting for the off-diagonal hyperfine interaction,
\begin{align}\ 
    A &= \frac{2}{3}(\nu_{3/2} - \nu'_{1/2}) \nonumber \\   
    &= \frac{2}{3}(\nu_{3/2} - \nu_{1/2} + \Delta\nu_{1/2}^{(2)}) \nonumber \\
    &= \frac{2}{3}\Delta\nu_{\mathrm{hfs}} + \frac{2}{3}\Delta\nu_{1/2}^{(2)} \label{eq:A'1} \\
      &= A_{\mathrm{eff}} + \frac{2}{3}\Delta\nu_{1/2}^{(2)} \nonumber
\end{align}
and so the  second-order correction to the $F=I$ level can be recovered from $A$ and $ A_\mathrm{eff}$ (the experimentally-determined value),
as is done in Section \ref{sec:OffDiag}. Equation (\ref{eq:Delta E}) is  the equality from (\ref{eq:A'1}) rearranged. 

\section{\label{app:Uncertainties} UNCERTAINTY EVALUATIONS } 
The mass shift and field shift parameters are not observable quantities and so no direct comparison can be made between computational values and experiment. Accordingly, an alternative method is used to determine the uncertainties of our calculated parameters. The method used here follows the method we used in Ref.~\cite{Schelfhout2021(2)} which is based on the suggestion that the desired value should be monitored as the computations are expanded in some systematic way \cite{Froese_Fischer2016}. To this end, a representative isotope for each element is chosen and the isotope shift parameters are calculated a number of times as the active space is extended. For Hg the representative is chosen to be ${}^{198}$Hg and for Cd it is ${}^{114}$Cd. We consider adding correlation layers separately to increasing the size of the available core and add the uncertainties estimated from these approaches in quadrature (i.e. assuming they are independent standard deviations).

\begin{figure*}[htbp]
    \centering
    \includegraphics[width=0.94\textwidth]{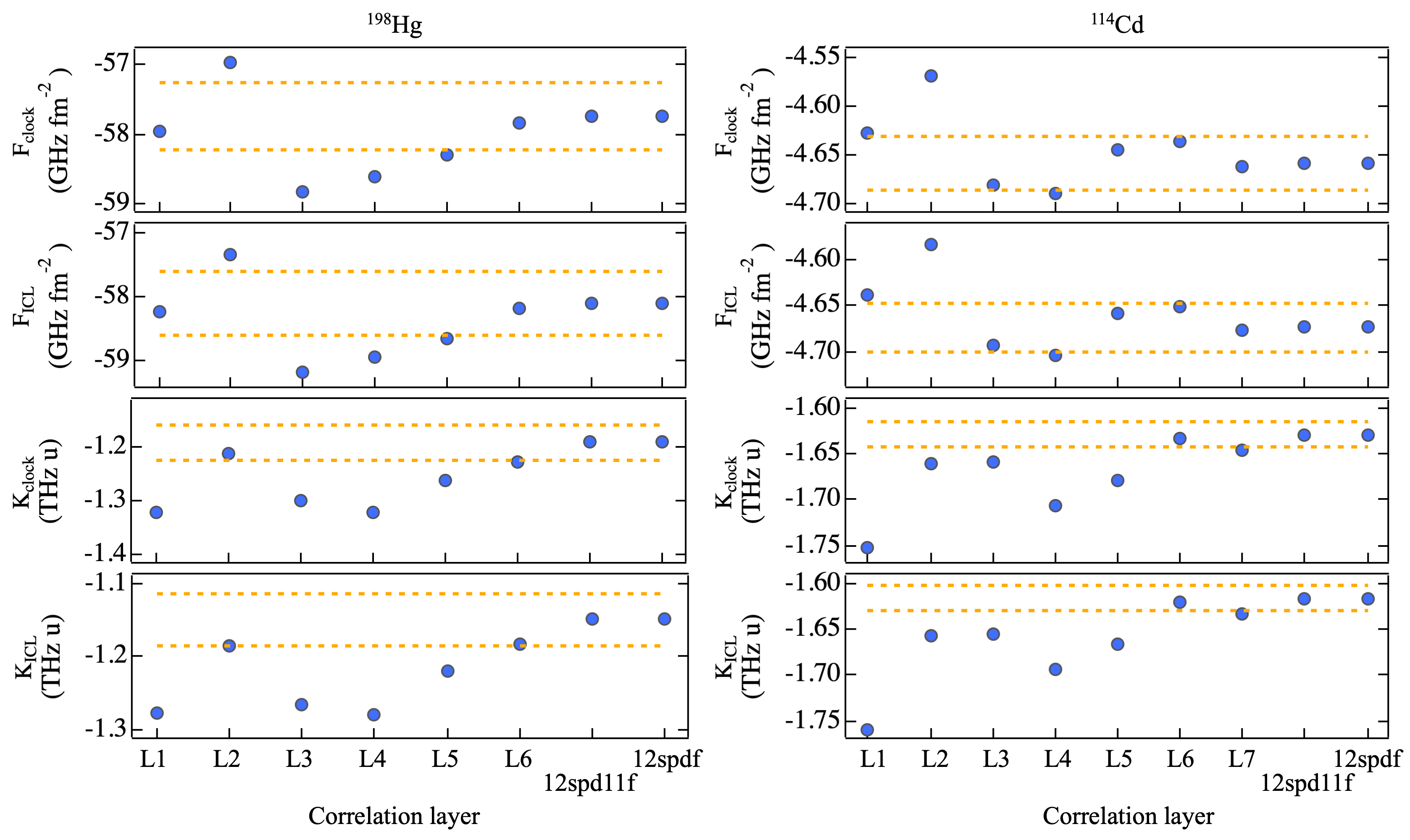} 
    \caption{ Sequences of mass shift and field shift parameters as the active space is expanded by adding correlation layers (denoted $L$). Uncertainty estimates (dotted lines) are presented around the final value.  The left panel is for $^{198}$Hg and the right panel is for  $^{114}$Cd. From top to bottom the plots show $F_{\mathrm{clock}}$, $F_{\mathrm{ICL}}$, $K_{\mathrm{clock}}$ and $K_{\mathrm{ICL}}$.   }
    \label{fig:convergence_up}
\end{figure*}

For the  monitoring of convergence as correlation layers are added, the relativistic configuration interaction (\textsc{rci}) routine is run after each layer is built with MCDHF calculations.  This is in contrast to our usual approach where \textsc{rci} is only run after adding  the final (partial) layer.   This way, the values seen in Fig.~\ref{fig:convergence_up} are consistent with $K$ and $F$ values presented earlier.  The sequences of the mass shift and field shift parameters for Hg and Cd are presented in Fig.~\ref{fig:convergence_up}.  The evaluated uncertainties are represented by the dotted lines, and are also listed in Table \ref{tab:convergence_up}.
The uncertainty is estimated to be the absolute difference between values
obtained from the two largest correlation layers that are complete layers.  This corresponds to labels L5 and L6 for Hg, and L6 and L7 for Cd in Fig~\ref{fig:convergence_up}, since the final two correlation layers are incomplete layers (the last full correlation layer is $12sp11d10f$).  This means of estimating the uncertainty may be  conservative given that in many cases the last three $K$ and $F$ values in the sequence have a standard deviation less  than this. However, the $13s$ and $13p$ orbitals are missing from the final layers, which are expected to have some influence (difficulties were encountered with the \textsc{grasp2018} computations when extended to $13sp$).

\begin{table}[htbp]
    \centering
    \caption{Estimates for uncertainties in mass and field shift parameters for Hg and Cd due to convergence of values when adding correlation layers.}
     \begin{ruledtabular}
    \begin{tabular}{ccccc}   
          Isotope & $\Delta F_{\mathrm{clock}}$ & $\Delta F_{\mathrm{ICL}}$ & $\Delta K_{\mathrm{clock}}$ & $\Delta K_{\mathrm{ICL}}$ \\ 
        & $(\mathrm{GHz\,fm^{-2}})$ & $(\mathrm{GHz\,fm^{-2}})$ & $(\mathrm{THz\,u})$ &  $(\mathrm{THz\,u})$ \\ \hline
        ${}^{198}$Hg & 0.48 & 0.49 & 0.034 & 0.036\\
        ${}^{114}$Cd & 0.027 & 0.027 & 0.014 & 0.014\\ 
    \end{tabular}
     \end{ruledtabular}
    \label{tab:convergence_up}
\end{table}

\begin{figure*}[htbp]
    \centering
    \includegraphics[width=0.94\textwidth]{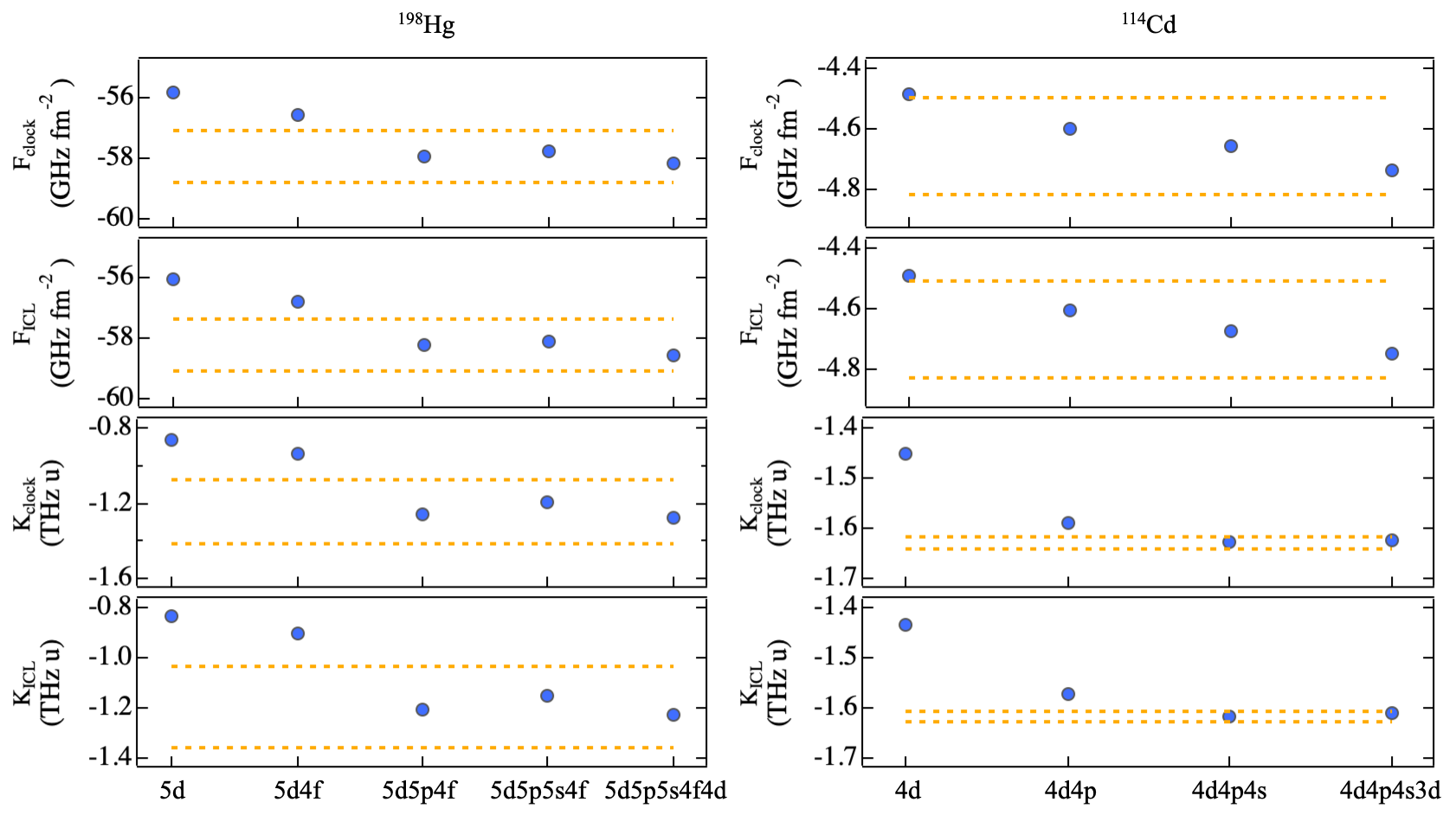} 
    \caption{Sequences of mass shift and field shift parameters as the active space is expanded by extending the available core. Uncertainty estimates (dotted lines) are presented around the penultimate value. The left panel is for $^{198}$Hg and the right panel  is for $^{114}$Cd. From top to bottom the plots show $F_{\mathrm{clock}}$, $F_{\mathrm{ICL}}$, $K_{\mathrm{clock}}$ and $K_{\mathrm{ICL}}$.}
    \label{fig:convergence_down}
\end{figure*}

For the monitoring of convergence as the active core is expanded, the method used to compute the main MCDHF-CI results is repeated several times with the input files to \textsc{rcsfgenerate} altered depending on the set of active core orbitals in consideration (\textsc{rcsfgenerate} generates the configuration state functions). The sequence is taken to one core orbital deeper than is used for the majority of calculations conducted in this paper. 
The uncertainty is taken to be  twice the absolute difference between the values for the two deepest  core extensions. 
The sequences of the mass shift and field shift parameters for Hg and Cd as substitutions to the core deepens are presented in Fig.~\ref{fig:convergence_down}. Again the 1-$\sigma$ uncertainty bounds are shown as dotted lines, but this time centered on the penultimate value. The estimated uncertainties are also reported in Table \ref{tab:convergence_down}.

\begin{table}[h] 
    \centering
    \caption{Estimates for uncertainties in mass and field shift parameters for Hg and Cd due to convergence of values when expanding the available core.}
    \begin{ruledtabular}
    \begin{tabular}{ccccc} 
        Isotope & $\Delta F_{\mathrm{clock}}$ & $\Delta F_{\mathrm{ICL}}$ & $\Delta K_{\mathrm{clock}}$ & $\Delta K_{\mathrm{ICL}}$ \\ 
        & $(\mathrm{GHz\,fm^{-2}})$ & $(\mathrm{GHz\,fm^{-2}})$ & $(\mathrm{THz\,u})$ &  $(\mathrm{THz\,u})$ \\ \hline
        ${}^{198}$Hg & 0.86 & 0.87 & 0.17 & 0.16\\
        ${}^{114}$Cd & 0.16 & 0.16 & 0.013 & 0.010\\   
    \end{tabular}
    \end{ruledtabular}
    \label{tab:convergence_down}
\end{table}

\bibliography{Refs2021}

\begin{thebibliography}{77}
\bibitem [{com()}]{comment1}%
  \BibitemOpen
  \href@noop {} {}\bibinfo {note} {The second term in Eq. 7 of this reference
  is not required for the calculations as the off-diagonal $B$ term is zero. We
  also note that the equality in Eq. 8 of the same reference should be
  proportional to.}\BibitemShut {Stop}%
\end{thebibliography}

\end{document}